\documentclass[fleqn,usenatbib]{mnras}

\usepackage{newtxtext,newtxmath}

\usepackage[T1]{fontenc}
\usepackage{ae,aecompl}
\usepackage{color}

\usepackage{graphicx}	
\usepackage{amsmath}	
\usepackage{amssymb}	
\usepackage{textgreek}
\usepackage{enumitem}
\usepackage{hyperref}

\title[Machine Learning Subhalos]{Machine Learning the Fates of Dark Matter Subhalos: A Fuzzy Crystal Ball}

\author[A. Petulante et al.]{
Abigail Petulante,$^{1}$\thanks{E-mail: abigail.petulante@vanderbilt.edu}
Andreas A. Berlind,$^{1}$
J. Kelly Holley-Bockelmann,$^{1,2}$ and
\newauthor Manodeep Sinha$^{1,3,4}$
\\
$^{1}$Department of Physics and Astronomy, Vanderbilt University, 2301 Vanderbilt Place, Nashville, TN 37235, USA\\
$^{2}$Department of Physics, Fisk University, 1000 17th Ave. N, Nashville, TN 37208, USA\\
$^{3}$Centre for Astrophysics and Supercomputing, Swinburne University of Technology, Hawthorn, Victoria 3122, Australia\\
$^{4}$7ARC Centre of Excellence for All Sky Astrophysics in 3 Dimensions (ASTRO 3D)
}

\date{Accepted XXX. Received YYY; in original form ZZZ}

\pubyear{2020}

\begin{document}
\label{firstpage}
\pagerange{\pageref{firstpage}--\pageref{lastpage}}
\maketitle

\begin{abstract}
The evolution of a dark matter halo in a dark matter only simulation is governed purely by Newtonian gravity, making a clean testbed to determine what halo properties drive its fate. Using machine learning, we predict the survival, mass loss, final position, and merging time of subhalos within a cosmological N-body simulation, focusing on what instantaneous initial features of the halo, interaction, and environment matter most. Survival is well predicted, with our model achieving 96.5\% accuracy using only 3 model inputs from the initial interaction. However, the mass loss, final location, and merging times are much more stochastic processes, with significant margins of error between the true and predicted quantities for much of our sample. The redshift, impact angle, relative velocity, and the masses of the host and subhalo are the only relevant initial inputs for determining subhalo evolution. In general, subhalos that enter their hosts at a mid-range of redshifts (typically z = 0.67-0.43) are the most challenging to make predictions for, across all of our final outcomes. Subhalo orbits that come in more perpendicular to the host are also easier to predict, except for in the case of predicting disruption, where the opposite appears to be true. We conclude that the detailed evolution of individual subhalos within N-body simulations is quite difficult to predict, pointing to a stochasticity in the merging process. We discuss implications for both simulations and observations.
\end{abstract}




\section{Introduction}
\label{sec:introduction}

According to the standard \textLambda CDM model of cosmology, dark matter structures in the universe form hierarchically through a series of mergers, with larger halos continuously growing from the accretion of smaller halos. Once independent halos themselves, these "subhalos" sink to the center of their "host" halos, losing mass to their hosts along their orbits due to tidal effects and dynamical friction, a process which has been studied in detail \citep{Tormen1998, Weinberg1989, Bosch1999, Hayashi2003, Taffoni2003, Gan2010, VandenBosch2017}. A significant number of such subhalos retain some of their mass, surviving as substructures within their hosts today. The study of these substructures has been fundamental to our understanding of many areas of astrophysics, from large-scale structure \citep{Zentner2003, Knebe2004, Zentner2004, Watson2011} to the formation and evolution of galaxies \citep{Hayashi2009, Kazantzidis2009, Simha2016}, which rely on both accurate final subhalo populations and the evolution of these populations \citep{Diemand2007, Giocoli2007}. 

Theoretical models of galaxy formation and evolution are commonly put to the test through analytic frameworks of subhalo evolution  \citep{Taylor2003, Zentner2004, VanDenBosch2005, Penarrubia2005, Jiang2016}. These techniques typically rely on merger trees constructed from N-body simulations, along with analytic treatments of the physical processes that cause their evolution, which allow for an in-depth study of the individual effects of dynamical friction, tidal stripping, and tidal heating. Semi-analytic models often perform quite well, producing galaxy populations that match hydrodynamic simulations \citep{Croton2006, DeLucia2007, Somerville2008, Henriques2010, Hirschmann2012, Mitchell2018}. Other works have focused on developing analytical models that reproduce specific properties of subhalos, such as disruption and mass loss rates, spatial distributions within host halos, and merging timescales, to better understand our assumptions about these driving physical processes.

Although there has been significant focus on constructing semi-analytic models of subhalo evolution, the {\em reliability} of tuning these models to N-body simulations has remained relatively unexplored. N-body simulations do produce consistent subhalo mass functions, the  evolution of which has been thoroughly studied \citep{Gao2004, Onions2012, Jiang2016a, Chua2016} even down to 10\textsuperscript{7} solar masses \citep{Munshi2018}. However, though the ensemble of subhalos is well-characterized, it is not clear that the evolution of an {\em individual} subhalo within these simulations is a truly deterministic process. Subhalo evolution models that are tuned to N-body simulations \citep{Penarrubia2005, Gan2010, Hiroshima2018} reflect the noise and uncertainty within the simulation. It may be that subhalo evolution within N-body simulations is somewhat stochastic, such that nearly identical interactions evolve differently, adding unforeseen complications when using these models.

In this work, we use halo merger tree data generated from the dark matter only simulation VISHNU, described in \citet{Johnson2019}, to quantify the evolution and fate of subhalos. We attempt to predict the final subhalo state using initial conditions at the time the subhalo enters its host. Using a large range of physically-motivated quantities at the time of a subhalo's entry, we train machine learning algorithms to predict final properties for the subhalo, in the hopes of investigating to what degree its fate is determined by these parameters and to what degree the interaction is stochastic and cannot be predicted. 
If the amount of mass loss, for example, is deterministic, a machine learning algorithm should be able to successfully map subhalo initial conditions to its final mass. On the other hand, if there is a level of stochasticity in subhalo fate, there will remain large prediction errors, even when using a complete set of inputs that describe its initial state.

Machine learning has emerged as a powerful tool in astrophysics with a variety of applications, such as galaxy classification \citep{Barchi2020, Nolte2019GalaxyData}, exoplanet detection \citep{Schanche2019}, and gravitational wave noise removal \citep{Cavagli2018}, and has recently been used with cosmological simulations to predict galaxy properties from halo properties \citep{Kamdar2016}, populate halos with galaxies \citep{Agarwal2017, Jo2019}, connect initial conditions to final halos \citep{Lucie-Smith2018}, and predict the halo masses of galaxies \citep{Calderon2019PredictionApproach} and clusters \citep{Ntampaka2015}. The ability of machine learning models to approximate any function with a large set of parameters provides a useful means of revealing complex correlations when a direct analytic function cannot be found. Notably, \citet{Nadler2017} recently used machine learning to predict the survival or disruption of subhalos in a hydrodynamic simulation, using the initial conditions of their counterparts in a dark matter only simulation. They were quite successful, accurately predicting the results of 85\% of their test set of subhalos. Works like this are encouraging that machine learning can be used to fit these complicated interactions. 

Here, we focus on dark matter only simulations and aim to predict not only survival, but more quantitative metrics such as the amount of mass loss, the final position, and the time to final merger for a subhalo. By using a comprehensive set of model inputs that describe the physical state of the subhalo to make these predictions, we hope to reveal what properties of the subhalo are the most closely tied to -- and thus what physical processes most strongly drive -- this evolution. While we expect these additional quantities to be more difficult to predict, even in a dark-matter only simulation with simpler physics, than a binary prediction for disruption, the ability or inability of machine learning models to make predictions in the first place can inform us about the determinism of a model. Predictions that are not successful, despite having a complete set of physical descriptors available to them, may indicate that subhalos in N-body simulations do not evolve in a straightforward, predictable way. This could be due to a number of factors, ranging from resolution effects, to halo catalog or merger tree errors, to an inherent chaotic nature of these interactions.

The topic of subhalo disruption has a particularly rich body of work, which will guide us in selecting our initial suite of halo properties to use as model inputs. Accretion redshift has repeatedly been found to be overwhelmingly important in determining the survivability of subhalos, with the majority of surviving subhalos being accreted more recently than z=1 \citep{Ghigna2000, Diemand2004, Gao2004, Zentner2004, Penarrubia2005, Diemand2007}. The abundance of subhalos is also found to be lower for host halos of fixed mass that have higher concentrations. Because higher concentration halos on average form earlier, they have less surviving substructure because their substructure is accreted earlier and spends more time orbiting inside the host \citep{Gao2004, Giocoli2009, Gao2011}. Trends with the orbits of subhalos find that many subhalos that are destroyed do not complete even one pericenter passage, but many subhalos that do survive have completed more than one pericenter passage. However, it has also been found that surviving subhalos tend to have more eccentric orbits \citep{Klimentowski2010} . Slower subhalos with low orbital energies are preferentially destroyed, resulting in a positive velocity bias in the subhalo distribution within clusters \citep{Diemand2004}. The fraction of surviving subhalos can be well modeled as a function of the subhalo-to-host mass ratio \citep{Tormen1998}, and subhalos with larger mass ratios have been found to more rapidly disrupt \citep{Tormen1998}. It has also been found that there is a weak dependence of the subhalo disruption rate on the mass of the host halo \citep{Gill2004}.

As subhalos typically disrupt after losing a significant fraction of their mass \citep{Taylor2005}, we expect many of the properties that determine subhalo survival to also be of significant importance to subhalo mass loss. For populations of subhalos across a wide variety of host halos, subhalo mass loss does not appear to strongly depend on the mass of the host halo \citep{Gao2004}. In a static host potential, the eccentricity of the subhalo orbit and the subhalo concentration are the dominant determinants of mass loss, with subhalos losing a significant portion of their mass during each pericenter passage, resulting in more radial orbits losing mass more quickly \citep{Taylor2003}. Average mass loss rates of subhalos using only the redshift and subhalo-to-host mass ratio appear to have good agreement with subhalo mass functions from simulations \citep{VanDenBosch2005}. Additionally, many analytical models of subhalo positions and internal structure at each timestep have been created to model subhalo evolution \citep{Taylor2001, Hayashi2003, Kampakoglou2007, Gan2010, Han2016}. Although these works give deeper insights to the relative importance of the physical processes at work on these subhalos, our machine learning models do not explicitly model the evolution over time of our subhalos, so the properties used by those works are less relevant here.

The final distributions of subhalos within their hosts have also been closely studied. Radial distributions of subhalos within their hosts do not appear to depend on host halo mass or redshift, but do depend on the subhalo-to-host mass ratio \citep{Angulo2009}. Subhalos that merge with the central parts of their host halos also tend to have larger subhalo-to-host mass ratios than those merging with the outer parts of the host halo, which may suggest that mass ratio can help predict the final location of a subhalo \citep{Nipoti2018}. Distributions of subhalo spins show lower spins closer to the host center, suggesting that lower spin halos may have more success at surviving to z=0 when orbiting at small fractions of the host radius \citep{Reed2005}. If this is due to higher spin subhalos being more susceptible to tidal stripping, this trend could also be important for our other predicted quantities. 

Analytic predictions of merging timescales for subhalos have been found by a number of previous works. A function to determine the time until satellite removal after accretion can be successfully fit using only the subhalo-to-host mass ratio \citep{Wetzel2010}. The effects of dynamical friction can be accurately modeled to determine galaxy merging timescales, using the subhalo-to-host mass ratio, the circularity and energy of the subhalo orbit, the virial radius of the host halo, and the dynamical time at the host halo's virial radius \citep{Boylan-Kolchin2007, Jiang2007, McCavana2012}. Although the dependencies on these parameters are different in these different works, the parameters dominating the merging timescale remain the same.

In Section~\ref{sec:data}, we describe the simulation and input data. In Section~\ref{sec:ML Methods}, we cover the machine learning methods we use to create our predictive models. In Section~\ref{sec: param selection}, we discuss the parameter selection methods we use to gain intuition and decide on which parameters are the most important for each model to make predictions. In the rest of Section~\ref{sec:results}, we share the results of our models for predicting each of our outcomes, including their performance and which parameters were needed as inputs to the model. Finally, in Section~\ref{sec:Conclusion} we discuss implications of our results for both observation and theory.

\section{Description of the Data}
\label{sec:data}
Our analysis makes use of VISHNU, a cosmological N-body simulation with 1000 snapshots for exquisite time resolution; no snapshot is separated by more than 3.2 $\times$ 10\textsuperscript{7} years. VISHNU contains 1680\textsuperscript{3} dark matter particles of mass \textit{m\textsubscript{p}} = 3.215 $\times$ 10\textsuperscript{7}h\textsuperscript{-1}M\textsubscript{\(\odot\)} in a box of size 130 h\textsuperscript{-1}Mpc and uses WMAP-1 cosmology \citep{Spergel2003}; $\Omega$\textsubscript{m} = 0.25, $\Omega$\textsubscript{$\Lambda$} = 0.75, $\Omega$\textsubscript{b} = 0.04, $\sigma$\textsubscript{8} = 0.8, \textit{n\textsubscript{s}} = 1.0, \textit{h} = 0.7). The initial positions and velocities of the particles at redshift z = 599 were then determined using the 2LPT code \citep{Scoccimarro1998}. The simulation was evolved to z = 0 using the GADGET-2 N-body TreeSPH code \citep{Springel2005}, adopting
a force resolution of 2.2 h\textsuperscript{-1}kpc. The ROCKSTAR halo finder was used to identify halos and subhalos \citep{Behroozi2013}, adopting a spherical overdensity halo definition with a threshold density equal to 200 times the background density of the universe. We denote the mass and radius of such halos with M\textsubscript{200b} and R\textsubscript{200b}. Finally, merger trees were constructed using the code \texttt{Consistent-Trees} \citep{Behroozi2013a}. 

Starting with halos at \textit{z}=0, we use the merger trees to identify the most massive progenitors of all host halos within the simulation, and track the subhalos within. These subhalos are allowed to host further substructure but cannot, at any point during their infall, become sub-substructure themselves. To help mitigate resolution uncertainties, we select only subhalos with a minimum of 1000 particles (total mass 3.215 $\times$ 10\textsuperscript{10}h\textsuperscript{-1} M\textsubscript{\(\odot\)}) at their time of accretion. We define the accretion time as the last snapshot before a subhalo enters its host. This mass cut reduces our sample from over 1,250,000 to 121,343 subhalos.

In addition to this resolution cut, some interactions were removed due to their unphysical behavior, likely as a result of errors in the merger tree generation or halo finder. Specifically, 1474 subhalos were removed that more than tripled their mass during infall, likely due to swapping identities with another halo. In addition, 327 subhalos were removed because their initial mass was larger than that of their supposed hosts. Taken together, these cuts culled about 1.5\% of halos, leaving 119,543 subhalo-halo interactions in our final sample.

Once a subhalo has been accreted by a host, it must have one of two fates: disrupt within the host (\textit{merge}), or remain a bound, identified subhalo within that host until today (\textit{survive}). We define the merger time as the last snapshot at which a subhalo is identified as its own entity. Following uncertainties in subhalo mass loss shown by \citet{Bosch2018}, we also consider a subhalo to be merged when it has lost more than 90\% of its mass, provided it remains underneath this threshold for the remainder of the simulation. This cut changes the fates of 21,929 subhalos, around 18\% of our total sample, but ensures a more consistent definition of merging that is less sensitive to resolution errors.

\begin{figure}
	\includegraphics[width=\columnwidth]{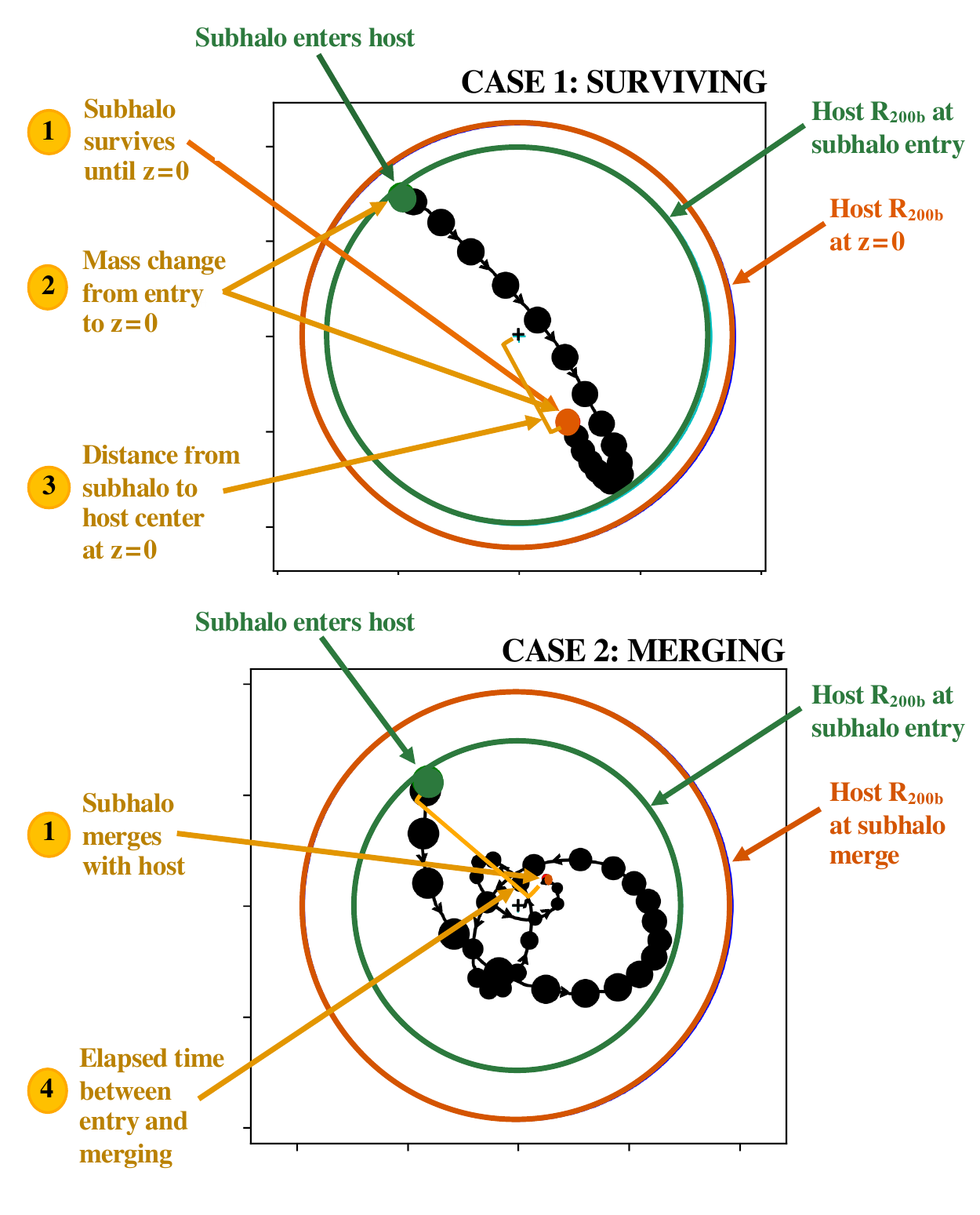}
	\vspace{-15pt}
    \caption{An example of a surviving (top) and merging (bottom) interaction between a subhalo and host halo. The large circles show the radii of the host halos at the beginning and end of the interaction. The orbits of the subhalos are shown with the series of filled circles, plotted at each eighth timestep, and with a point size corresponding to subhalo mass along the orbit. The green point and circle show initial quantities, at the timestep right before the subhalo enters its host. The orange point and circle show final quantities, at either the timestep right before the subhalo dissolves in the merging case, or at the final timestep in the simulation in the surviving case. Predicted quantities (gold) are labeled and numbered in the order we will present them throughout the paper.}
    \label{fig:explanatory_figures}
\end{figure}

\begin{figure*}
	\includegraphics[width=\textwidth]{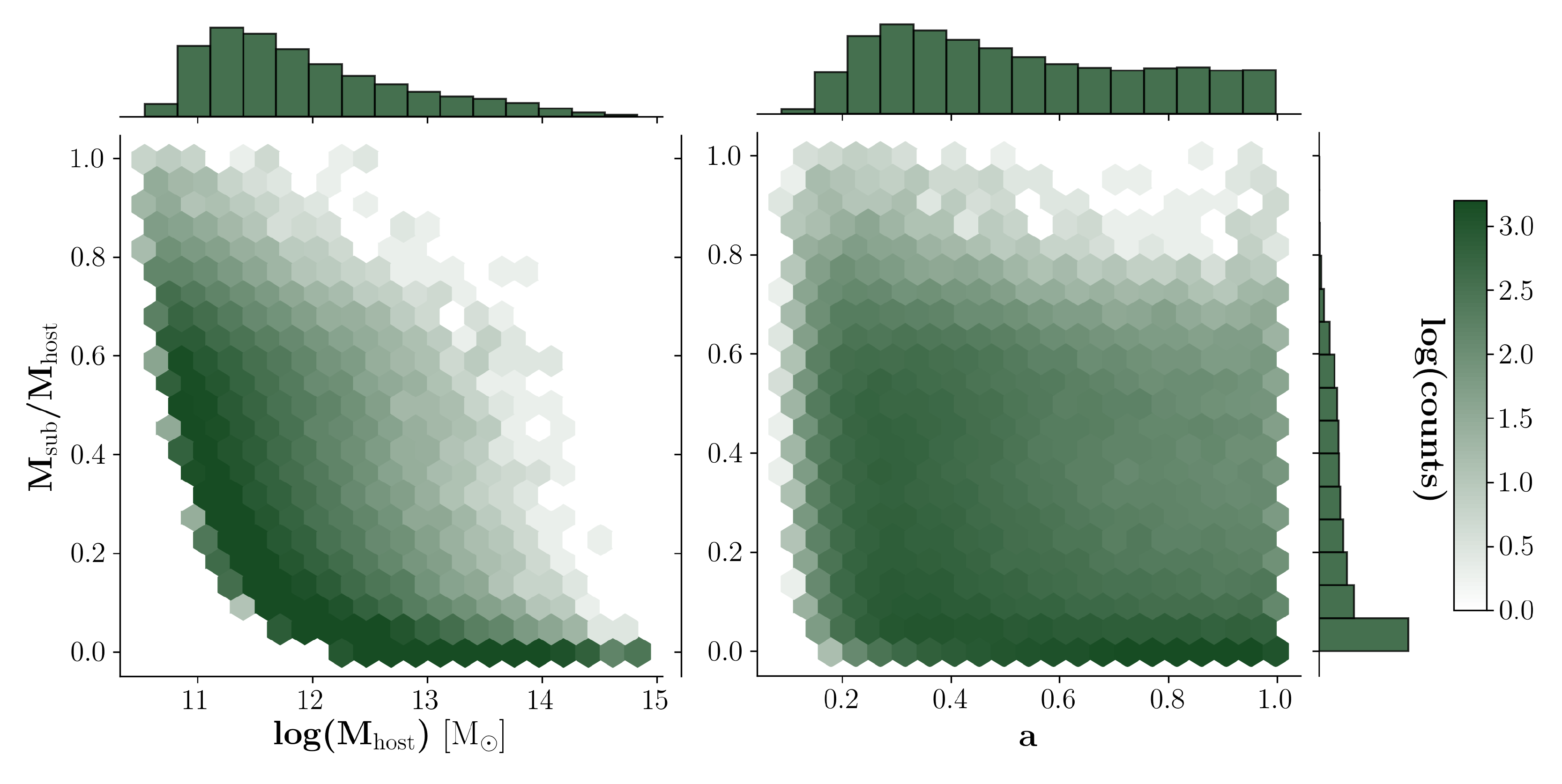}
	\vspace{-20pt}
    \caption{Demographics of our sample of interactions. The left panel shows the distribution of mergers as a function of host mass and mass ratio, defined at the time of accretion. The cosmological scale factor of accretion versus mass ratio is shown on the right. The color represents number of mergers in each hexagonal pixel on a logarithmic scale, as denoted on the far right. Histograms show the one-dimensional distribution of the variable on the corresponding axis. Most mergers occur at lower mass ratios and for smaller host halos, but the spaces are still well-spanned over a range of interactions in both panels. The chosen cut of 1000 particles (M = 3.215 $\times$ 10\textsuperscript{10}h\textsuperscript{-1} M\textsubscript{\(\odot\)}) for subhalos upon entry is clearly shown in the left panel. Imposing this cut and requiring that host halos be larger than their subhalos means that, for very small host halos, the only subhalos within our dataset are those with masses more similar to their hosts.}
    \label{fig:combined_distributions_logBin}
\end{figure*}

Figure~\ref{fig:explanatory_figures} illustrates the two possible subhalo fates using actual examples from our data set. The top panel shows a surviving interaction, where the subhalo orbits for some time, losing mass but not dissolving before z=0, while the bottom panel demonstrates a merger. In all, 76,442 (64\%) are mergers and 43,101 (36\%) survive. The four quantities we predict are shown in yellow and numbered by the order that they will be presented in Section~\ref{sec:results}. Note that the binary fate, survival or merger, uses the whole sample, but mass loss and final position considers only surviving halos and merging time applies only to those halos which merge. 

Distributions of our final sample with respect to host masses, mass ratios, and the scale factor of the time of entry are shown in Figure~\ref{fig:combined_distributions_logBin}. The most common interactions are those of unequal masses that occurred more recently. However, our sample also spans the space of more equal mass and higher redshift interactions, with hundreds of interactions shown in many of the bins in Figure~\ref{fig:combined_distributions_logBin}. The effects of our chosen particle cut can also be clearly seen in Figure~\ref{fig:combined_distributions_logBin}. Because subhalos must have 1000 particles at their time of entry, this results in a minimum initial mass of 3.215 $\times$ 10\textsuperscript{10}h\textsuperscript{-1} M\textsubscript{\(\odot\)} for both the subhalo and host halo, given that a host halo must also be at least as large as its subhalo. In the left panel of Figure~\ref{fig:combined_distributions_logBin}, this results in an area of no data with low mass hosts and unequal masses, because most subhalos of low mass hosts are too low mass to be included in our sample. We also do not have many interactions between very large hosts halos and similarly large subhalos. This is because there are relatively few massive halos in the simulation, so interactions between them are expected to be very rare.
   
Here we describe our set of physically-motivated parameters to characterize an interaction.
Our parameters comprise four categories: 1) Global interaction parameters give information about the interaction that is not specific to the particular system; 2) Internal halo parameters that are properties of the individual halos themselves and describe their size, shape, or structure; 3) Orbital parameters provide information about the initial trajectory of the subhalo's infall path; and 4) Environmental parameters describe the influence of larger scale environment around the subhalo and its host. In total, this yields 26  parameters, which we list and define here. All of these quantities are measured at the time of accretion, which is defined as the last snapshot before a subhalo enters its host.
    
\vskip 0.1in
    \noindent\textbf{Global Interaction Parameters:}
    \begin{itemize} [leftmargin=.4cm,topsep=0pt]
        \item \textbf{a}: the scale factor of the universe.
        \item \textbf{q}: the ratio of subhalo to host halo masses.
    \end{itemize}
\vskip 0.1in
    \noindent\textbf{Internal Halo Parameters:}
    \begin{itemize}[leftmargin=.4cm,topsep=0pt]
        \item \textbf{M\textsubscript{sub}}: Mass of the subhalo (just before it becomes a subhalo) at 200 times the background mass density of the universe,  M\textsubscript{200b}. This is defined using a spherical overdensity, including only those particles that were assigned to the subhalo and the sub-substructure within it.
        \item \textbf{M\textsubscript{host}}: M\textsubscript{200b} of the host halo, as described above. Note that this mass is calculated by including the particles of all substructure within the halo.
        \item \textbf{R\textsubscript{sub}}: Radius of the subhalo at the point where the mean subhalo density is 200 times the background mass density of the universe, R\textsubscript{200b}. As with M\textsubscript{200b} for the subhalo, this is defined using only particles belonging to the subhalo and its sub-substructure.
        \item \textbf{R\textsubscript{host}}: R\textsubscript{200b}, as described above, of the host halo. As with M\textsubscript{200b} for the host halo, this is defined using particles belonging to both the host halo and all of its substructures.
        \item \textbf{c\textsubscript{sub}}: the concentration of the subhalo, defined as R\textsubscript{200b}/R\textsubscript{s}, where R\textsubscript{s} is the scale radius of the subhalo.
        \item \textbf{c\textsubscript{host}}: the concentration of the subhalo, defined as R\textsubscript{200b}/R\textsubscript{s}, where R\textsubscript{s} is the scale radius of the host halo.
        \item \textbf{\textlambda\textsubscript{sub}}: Bullock spin parameter of the subhalo, defined as in \citet{Bullock2001}.
        \item \textbf{\textlambda\textsubscript{host}}: Bullock spin parameter of the host halo.
        \item \textbf{T\textsubscript{sub}}: the triaxiality parameter of the subhalo. Calculated from the definition given in \citet{Franx1991}:
        \begin{equation}
            T = \frac{1 - (b/a)^2}{1 - (c/a)^2}
        \end{equation}
        where b/a is the minor/major axis ratio and c/a is the intermediate/major axis ratio.
        \item \textbf{T\textsubscript{host}}:  the triaxiality parameter of the host halo, calculated as above.
        \item \textbf{max(M\textsubscript{subs,sub})}: M\textsubscript{200b} of the most massive sub-subhalo within the subhalo. 0 if subhalo has no sub-substrucutre. 
        \item \textbf{max(M\textsubscript{subs,host})}: M\textsubscript{200b} of the most massive subhalo already within the host halo at the time of the selected subhalo's entry. Does not include the selected subhalo. 0 if no other subhalos are present.
        \item \textbf{N\textsubscript{subs,sub}}: the total number of sub-subhalos within the subhalo.
        \item \textbf{N\textsubscript{subs,host}}:  the total number of subhalos within the host halo. Does not include the selected subhalo.
    \end{itemize}
\vskip 0.1in
    \noindent\textbf{Orbital Parameters:}
    \begin{itemize} [leftmargin=.4cm,topsep=0pt]
        \item \textbf{d\textsubscript{rel}}: distance between the centers of the subhalo and host halo.
        \item \boldmath{$v_\mathrm{rel}$}: magnitude of the relative velocity between subhalo and host halo, calculated in the reference frame of the subhalo.
        \item \boldmath$\epsilon$\unboldmath: eccentricity of subhalos initial orbit. Calculated as described in \citet{Wetzel2011}:
        \begin{equation}
            \epsilon = \sqrt{1 + \frac{2EL^2}{(GM\textsubscript{host}M\textsubscript{sub})^2\mu}}
        \end{equation}
        In this definition, $\epsilon$ = 1 is a perfectly elliptical orbit, and orbits that are initially unbound have eccentricities greater than 1.
        \item \boldmath$\phi$\unboldmath: impact angle of subhalos initial orbit. Calculated as $L_\textsubscript{total}/L_\textsubscript{max}$, the ratio between the total angular momentum of the subhalo orbit and the angular momentum of an orbit with the same velocity magnitude and orbital radius, but with the entire velocity component in the direction perpendicular to the direction of the host center. We refer to this as an impact angle because:
        \begin{equation}
            \frac{L\textsubscript{total}}{L\textsubscript{max}} = \frac{m(\vec{v} \times \vec{r})\textsubscript{total}}{m(\vec{v} \times \vec{r})\textsubscript{max}} = \frac{m v r cos(\theta)}{m v r} = cos(\theta)
        \end{equation}
        In this definition, $\phi$ = 1 is an orbit coming in perfectly perpendicular to the axis between the host and subhalo. We call this a grazing impact angle. Alternately, $\phi$ = 0 would be an orbit coming in perfectly along the axis between subhalo and host halo, which we call a plunging impact angle. We note that, while an orbit with $L_\textsubscript{max}$ would be instantaneously circular, this is different from the typical definition of circularity, $L_\textsubscript{total}/L_\textsubscript{circ}$, which compares the total angular momentum of the subhalo orbit and the angular momentum of a stable circular orbit with the same energy. As circularity is simply mathematically related to our definition of $\epsilon$, it would contain identical information.
    \end{itemize}
\vskip 0.1in    
    \noindent\textbf{Environmental Parameters:}
    \begin{itemize}[leftmargin=.4cm,topsep=0pt]
        \item \textbf{F\textsubscript{tid}}: Magnitude of the tidal force on the subhalo, approximated as the tidal force from the neighboring halo within d~= 4h\textsuperscript{-1}Mpc that contributes the most to the tidal force. Excludes the subhalo's own host.
        \begin{equation}
            F\textsubscript{tid} = \frac{M\textsubscript{neighbor}} {d\textsubscript{neighbor}\textsuperscript{3}}
        \end{equation}
        \item \boldmath$\rho$\unboldmath\textbf{\textsubscript{1Mpc}}: The density due to neighboring halos in a surrounding sphere with radius r~= 1h\textsuperscript{-1}Mpc from the subhalo center, not including the subhalo's own host halo.
        \begin{equation}
            \unboldmath{\rho\textsubscript{1Mpc}} = \sum_{i}{\frac{M\textsubscript{i}} {\frac{4}{3}\pi r\textsuperscript{3}} }
        \end{equation}
        \item \boldmath$\rho$\unboldmath\textbf{\textsubscript{2Mpc}}: The density due to other halos in a surrounding sphere with radius r~= 2h\textsuperscript{-1}Mpc from the subhalo center not including the subhalo's own host halo.
        \item \boldmath$\rho$\unboldmath\textbf{\textsubscript{4Mpc}}: The density due to other halos in a surrounding sphere with radius r~= 4h\textsuperscript{-1}Mpc from the subhalo center not including the subhalo's own host halo.
    \end{itemize}
\vskip 0.1in

We aim to use machine learning to predict the following for each subhalo:
    \begin{enumerate}[leftmargin=.4cm]
        \item [1.] \textbf{survival}: a (categorical, binary) indication of whether or not a subhalo survives until z=0. 0 or 1, depending on whether the subhalo exists above the required mass threshold at z=0 (survives, 1) or if the subhalo has fallen below the mass threshold at some time before z=0 (dissolves, 0). This mass threshold is defined as 10\% of the subhalos mass upon accreting into the host (M\textsubscript{sub}). Predicted for all subhalos.
        \item [2.] \textbf{M\textsubscript{sub,f}}: the M\textsubscript{200\textsubscript{b}} (as described above for M\textsubscript{sub}, calculated using only particles which belong to the subhalo) of the subhalo at z=0. Only predicted for surviving subhalos. When compared to the initial subhalo mass, this shows the amount of mass loss that the subhalo experiences.
        \item [3.] \textbf{d\textsubscript{rel,f}}: relative absolute total distance between subhalo and host halo centers at z=0, normalized by the radius of the host halo (R\textsubscript{host}, as described above) at z=0. Only predicted for surviving subhalos.
        \item [4.] \textbf{t\textsubscript{merge}}: the elapsed time between the accretion of a subhalo into the host and its dissolution within the host. Only predicted for dissolving subhalos.
    \end{enumerate}
\par

Machine learning, requires data to {\em train} and {\em test}; we randomly split our data into subsamples, with 80\% for training and 20\% to test. We scale and normalize the data using \texttt{StandardScaler} from the \texttt{scikit-learn preprocessing} package such that each quantity, X, if assumed Gaussian, is distributed with zero mean and unit variance: 
    \begin{equation}
        X\textsubscript{norm} = \frac{X - \mu}{\sigma}
    \end{equation} 
where \textmu{} is the mean of the unscaled data and \textsigma{} is the standard deviation. This scaling is necessary for many machine learning models, as large variations dynamic range over a set of observables can affect model accuracy.

\section{Machine Learning Methods}
\label{sec:ML Methods}
The machine learning algorithms we use come from the \texttt{scikit-learn} package for python. Subhalo survival is a classification problem, so we use a random forest algorithm. On the other hand, predictions of the amount of mass loss, the final position, and merging time are all classic regression problems, so we use the gradient boosting regressor algorithm. Although we refer the reader to the \texttt{\href{https://scikit-learn.org/stable/user_guide.html}{scikit-learn}} documentation for a full description of these algorithms, we briefly describe these methods below.

\subsection{Random Forest}
\label{sec:rf}
Random forest classifiers use an ensemble of decision trees to reach consensus on a prediction. These decision trees repeatedly split the data into bins based on the values of its input parameters, resulting in gradually smaller subsets of data belonging to each bin. The goal of the algorithm is to find bins that span a section of the input parameter space where almost all members of the bin have the same output value. Then, the assumption is that test data points with input parameters that fall in a certain bin will usually have the same output value as other members of that bin. In a random forest, many individual decision trees are trained on random subsets of the training dataset. Because random forests are a type of bagging - or bootstrap aggregating - method, the classification for an object is then the majority vote of all of the trees. This means that the trees that make up the ensemble are distinct, making an independent prediction for the classification of an object in the testing set.

There are several {\em hyperparameters} of the algorithm that we tune in order to get the best-fitting model. These hyperparameters are parameters of the model itself, that determine properties such as the complexity of the model or the way that it learns. Their values are fixed before the model is trained, and are not adjusted during the training of the model. The hyperparameters that we set for the random forest classifier are as follows. The \textit{n\_estimators} hyperparameter sets the number of estimators, in this case decision trees, used in the final consensus. Too few decision trees removes the power of using multiple trees. In general, using too many trees is not a concern, though it does increase the runtime of the algorithm.  The \textit{max\_depth} hyperparameter sets the maximum number of decisions in each tree. Effectively, this sets the maximum number of input parameters each decision can use, since each depth splits on one parameter. The \textit{max\_leaf\_nodes} hyperparameter sets the maximum number of nodes at a given depth. We note that, if this value is too small, the decision tree may split on the same parameter at multiple depths. We keep other hyperparameters at default values (see \texttt{scikit-learn} documentation).

One of the main advantages of random forest algorithms is that they are less prone to overfitting, especially compared to a single decision tree. Because each decision tree works with a subset of the data and considers a random group of parameters, the ensemble is more robust to unseen data. This is important in cases like ours in which a large number of training examples determine a small number of phenomena. Random forests are also useful in their ability to deal with correlations between input parameters, such as halo mass and concentration. Unfortunately, the results of a random forest are much less straightforward to interpret than that of single decision trees, and any given decision tree within the forest may be a very poor predictor. Nonetheless, random forest classifiers robustly rank the importance
of each input parameter, based on their frequency and proximity to the top of the decision trees within the forest. However, strong correlations between input parameters can make this ranking difficult to straightforwardly interpret as well, which we discuss in more detail in Section~\ref{sec: param selection}.

Subhalo survival is particularly amenable to binary classification; we assign 0 to a subhalo for dissolving before z=0, and a value of 1 for surviving until z=0. We train multiple models, using the same hyperparameters, on increasingly smaller subsets of input parameters to confirm the minimum number needed to make accurate predictions. To determine the order of parameter removal, and for presentation purposes in our figures, we use the custom parameter selection algorithm outlined in Section~\ref{sec: param selection} to determine the relative importance of our parameters.

\subsection{Gradient Boosting Regressors}
\label{sec:gbr} 
Gradient boosting regressors, like random forests, rely on an ensemble of decision trees to make predictions. However, unlike random forests, gradient boosting regressors construct trees that are dependent on the results of all the trees trained before them. New trees are fit to the errors of the current ensemble; the purpose of each new tree is to learn the errors of the current model and to iteratively hone in on the prediction.

As with the random forest classifier, several hyperparameters can be tuned to create the best model. The {\em learning rate} weights the significance of a new tree within the ensemble; small learning rates significantly increase the number of trees that need to be added to the model, but too large a learning rate may result in corrections that perpetually overshoot the prediction. As with the random forest classifier algorithm, we also tune the \textit{n\_estimators}, \textit{max\_depth} and \textit{max\_leaf\_nodes}. Other hyperparameters are kept as their default values.

The advantage of gradient boosting regressors is again the reduction of overfitting because it is an ensemble method. Additionally, given that we also expect relatively few parameters to be important in making our regression predictions, a model that is able to avoid selecting unimportant parameters is favorable. The main advantage that a gradient boosting ensemble has over a random forest is that the trees work together to make a prediction. Because each tree added to a gradient boosting model corrects errors from the previous iteration of the ensemble, data points that were difficult to make predictions for are preferentially corrected for in later trees, making a gradient boosting ensemble better at dealing with outliers. This is particularly important for our regression problems given the large ranges of outcomes for all of our predicted quantities. As with the random forest classifier, the gradient boosting regressor class in \texttt{scikit-learn} also contains a function to report relative feature importances. However, interpreting these results again presents difficulties due to strong correlations between parameters.

We use gradient boosting regressors to predict mass loss, final position, and merge time individually.  As with the survival classification problem, due to the difficulty of interpreting the reported feature importances of the algorithm, we use our custom algorithm, outlined in Section~\ref{sec: param selection} to determine the order and relative importance of each parameter.

\subsection{Model Training}
\label{model details}
When training any machine learning model, the choice of hyperparameters is critical for the model's ability to learn. So we begin by finding a set of hyperparameters for each of our models that leads to the best fit for our data. We do this by repeatedly creating models with different hyperparameters, and evaluating their performance using training and validation sets that are random subdivisions of our total training set, using an 80\%/20\% split. We then check if the model performs well without overfitting too strongly. From our complete dataset, these repeated divisions mean that 20\% of our data is used only for final testing, while 80\% of our data set is used to create different training and validation sets. The use of these validation sets allows us to fine tune the properties of the model so that it best fits the training set, while leaving the testing set untouched to ensure that the model performance at testing time accurately shows the model's ability to generalize to new data.

We begin this process by using \texttt{scikit-learn}'s \texttt{GridSearchCV} function, which accepts arrays of values for the desired hyperparameters to be tuned, then repeatedly trains the model on all combinations of the chosen hyperparameter values and evaluates each combination's performance. This function includes a \texttt{best\_params\_} attribute, which will return the combination of hyperparameters that leads to the best performance. We then, by inspection, fine-tune the hyperparameters within small ranges around this set. For example, we may begin with a depth hyperparameter of 5 for our model, then check if either raising the depth to 6 or lowering it to 4 with other hyperparameters held constant will give us improvement. We repeat this process for all hyperparameters, until we find a model that achieves high accuracy on both the training and validation sets, indicating that the model is well fit to the data, but with the smallest differences between training and validation accuracy, indicating that the model is minimally overfit. We take this extra step of fine-tuning the hyperparameters by hand to ensure that the model is well fit to our specific accuracy metrics for each prediction, which we motivate from physical quantities about the subhalo and host halo. These specific accuracy metrics that we use for each of our predicted quantities are described further in Section~\ref{sec:results}.

The best set of hyperparameters for each of our models can be found in Table~\ref{tab:hyper_table}. We note that, while we find these hyperparameters to create models that fit the data well, different choices of the hyperparameters can yield equally good models. Typically, slight changes to these hyperparameters did not lead to significant performance differences for our models. In the case of \textit{max\_leaf\_nodes}, we found that changing its value, even significantly, did not have strong effects on the model performance for any of our models, and thus we kept it fixed to its default value (\texttt{None}) for all of our models. However, large changes to the other hyperparameters can cause significant changes to the model's ability to fit the data, so using a selection of a well-fitting set of hyperparameters is imperative to getting the best results.

\begin{table}
	\centering
	\caption{The best fit hyperparameters for each of our models. Definitions of these hyperparameters and how they were selected are detailed in Section~\ref{sec:ML Methods}.}
	\label{tab:hyper_table}
	\begin{tabular}{c|cccc} 
		\hline
		  & Survival & Mass Loss & Position & Merge Time\\
		\hline
		method & RF & GBR & GBR & GBR\\
		\hline
		n\_estimators & 50 & 600 & 1800 & 500\\
		max\_depth & 7 & 3 & 6 & 5\\
		max\_leaf\_nodes & None & None &  None & None\\
		learning rate & N/A & .07 & .008 & .05 \\
		\hline
	\end{tabular}
\end{table}

\section{Results}
\label{sec:results} 

In the following subsections, we discuss our exploration of the parameter space and the performance of each of our machine learning models. In Section~\ref{sec: param selection}, we detail our parameter selection methods, where we find a preliminary order of the importance of our parameters for predicting our final quantities. Then, in the subsections that follow, we detail the  results of our machine learning models, including a discussion of our metrics for determining the accuracy of our predictions, how well each of the models perform, and which parameters were the most important for making the predictions for each model. In Section~\ref{sec:survival}, this is discussed in detail for predicting the survival quantity. In Section~\ref{sec:mass loss}, we discuss this in detail for the mass loss quantity. In Section~\ref{sec:position} we discuss the final position quantity, and in Section~\ref{sec:merge time} we discuss the merging time quantity. In Section~\ref{sec:interactions}, we investigate the frequency of subhalo interactions that we find within our sample and their potential effects on our predictions.

\subsection{Feature Selection}
\label{sec: param selection}

We select 26 features to describe each subhalo/host halo interaction. These parameters are selected to encompass information about the orbit, environment, and individual properties of both the host and subhalo. Although we begin with this large set of features for thoroughness, we expect that not all of them will be important for predicting our desired quantities. To determine which features most strongly affect the predicted quantities, we use a feature selection method to select four features from the complete set, for each predicted quantity, which are responsible for the most variation in that quantity. Because our set of features has strong correlations between several values, we also aim to use a selection method that minimizes correlations in the selected set. We emphasize that we do not remove any features using these methods. Instead, we use the order of the selected set to determine the order in which we will add features to our models and for display purposes in our figures. During training, all models still use all features, to ensure that no information is missed.

When building decision trees, higher ranked features are both the most important, and those that cause the most variance in the final quantity. Our aim, then, is to order the features by the amount of variance, while binning to remove correlations with other features; this leads to an independent ordering that appears to add information to the prediction most quickly.
We describe the feature selection method using the example of predicting the final mass of surviving subhalos. We begin by binning the data by each of our features, say impact angle, into bins with equal numbers of subhalos, and calculating the mean final mass in each bin. We then determine the range of these binned mean values as a measure of the strength of the correlation between final mass and impact angle. We adopt the feature with the largest range as the most important. This algorithm selects the subhalo radius as the most important feature for predicting the final subhalo mass. To select the next most important feature, we bin the data in two dimensions, where the one dimension is our adopted primary feature and for the second dimension we try each of the remaining features. In each two-dimensional bin we calculate the mean final mass. We can now measure the strength of correlation between final mass and each feature {\it at fixed} subhalo radius. We do that by calculating the range of mean final mass values across all bins of the secondary feature, while staying in the same bin of subhalo radius. Finally, we calculate the mean such range, averaging over all the bins of subhalo radius. We adopt the secondary feature with the largest mean range as the second most important feature. This algorithm selects the scale factor of the halo-subhalo interaction as the second most important feature for predicting the final subhalo mass. We continue with this process until we have extracted four features; the sample size does not permit further binning of the data beyond four dimensions without having too few bins to be useful.

The four most important features for each of the predicted quantities are shown in Table~\ref{tab:FS_table}. The numbers in parentheses represent the strength of correlation between the predicted quantity and each selected feature. This is determined by using the ranges, as described above, that were maximized to select the most important features. For ease of comparison, we normalize these ranges by the full range of the predicted quantity in the data. For example, in the case of predicting the final position of the subhalo, the range of mean final positions in bins of scale factor is 53.8\% of the total range of final positions. Furthermore, the mean range of final positions in bins of mass ratio, at fixed scale factor, is 21.4\% of the total range of final positions. The mean range of final positions in bins of impact angle, at fixed scale factor {\it and} mass ratio, is 17.7\% of the total range of final positions, and so forth. We note that in the case of mass loss, we convert to log space before reporting the normalized ranges.

\begin{table}
	\centering
	\caption{The ranking order of most important features for predicting each of the desired quantities. The second column displays the  normalized maximum variation resulting from binning that feature, where higher values indicate the feature is more strongly responsible for changes in the prediction outcome.}
	\label{tab:FS_table}
	\begin{tabular}{c|cccc} 
		\hline
		rank & Survival & Mass Loss & Position & Merge Time\\
		\hline
		1st & a (.997) & R\textsubscript{sub} (.243) & a (.538) & a (.283)\\
		2nd & $q$ (.254) & a (.132) & $q$ (.214) & $q$ (.127)\\
		3rd & $\phi$ (.145) & $\phi$ (.047) &  $\phi$ (.177) & $\phi$ (.091)\\
		4th & $v_\mathrm{rel}$ (.093) & M\textsubscript{host} (.042) & $v_\mathrm{rel}$ (.153) & $v_\mathrm{rel}$ (.058)\\
		\hline
		random & .032 & .013 &  .049 & .019\\
		\hline
	\end{tabular}
\end{table}

\begin{figure*}
	\includegraphics[width=\textwidth]{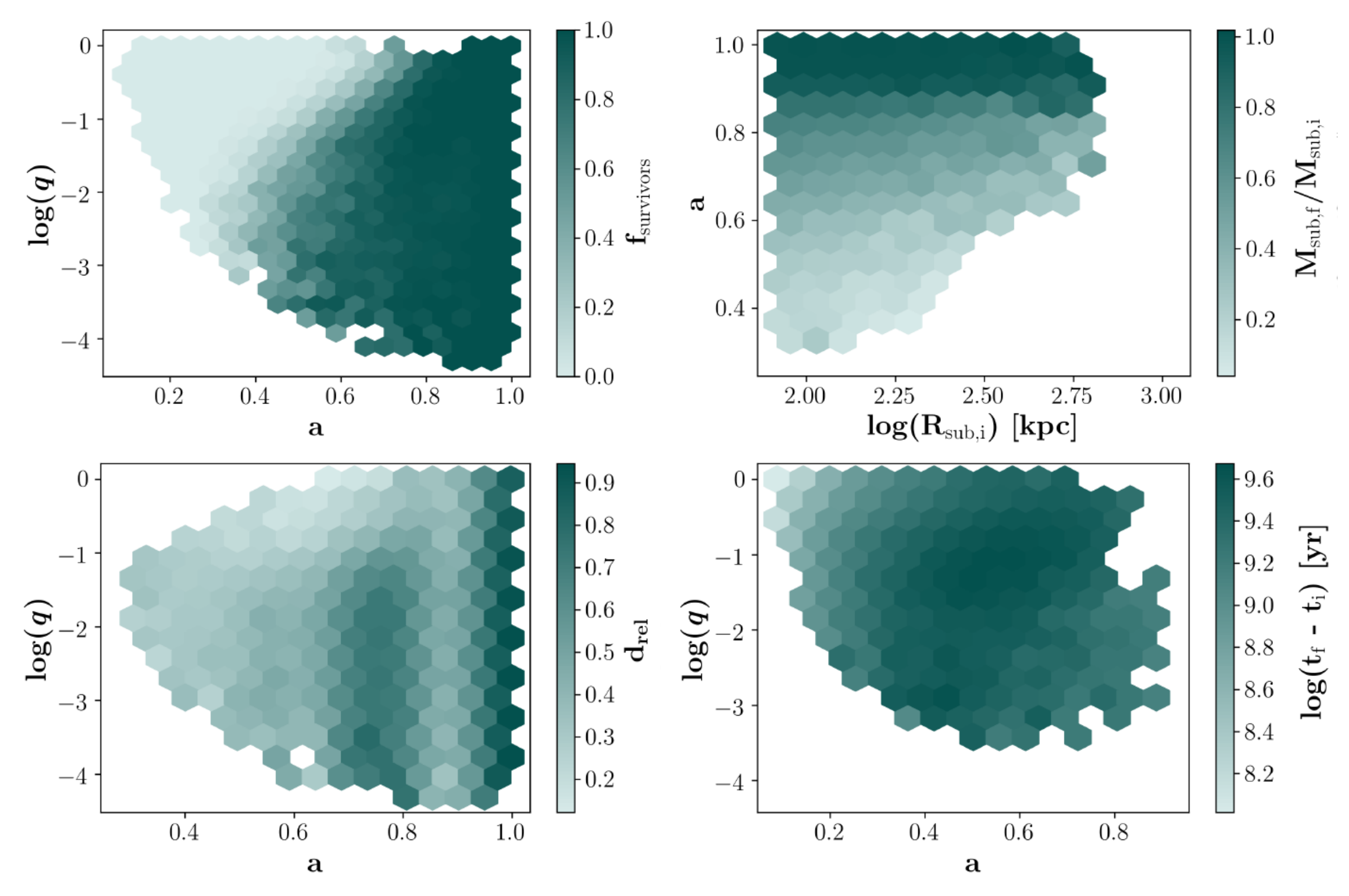}
	\vspace{-15pt}
    \caption{Distributions of the predicted quantities of interest with respect to the two parameters that are most responsible for each of their variations. In each panel, the parameter that causes the most variation is shown on the x-axis, and the parameter that causes the second most variation is shown on the y-axis. In addition, in each panel the colorbar shows the average value of the quantity of interest within a hexagonal bin. The top left panel shows the fraction of surviving subhalos. The top right panel shows the fraction of subhalo mass that remains for surviving subhalos. The bottom left panel shows the fractional distance of surviving subhalos from their host's center. The bottom right panel shows the elapsed time for a subhalo to merge. In each instance, some pattern of color striation can be seen to represent the importance of the two parameters shown. However, it is clear that the survival of a subhalo is by far the most drastically divided and well-defined by this two-dimensional space. }
    \label{fig:bestSpaces}
\end{figure*}

From Table~\ref{tab:FS_table}, we see that the same features appear repeatedly for all of our quantities. In particular, survival, final position, and merge time have the same four features, in the same relative order, chosen as most important for making their predictions. The initial scale factor of entry, ranked as most important for all of those quantities, is also ranked as second most important for predicting mass loss. The impact angle of the subhalos orbit also appears as important for predicting mass loss. These results are already encouraging, as many of these features are those that we were motivated to select because they were known to be important from previous works, as discussed in Section~\ref{sec:introduction}. It is interesting that, although the eccentricity of a subhalo has been shown to affect subhalo evolution in a number of ways \citep{Taylor2003, Boylan-Kolchin2007, Jiang2007, Klimentowski2010, McCavana2012}, the impact angle feature is chosen over the eccentricity, meaning that just the initial direction that the subhalo enters the host is more influential than the actual orbit that the subhalo is on. While these two parameters are fairly well correlated and contain similar information, there is a large amount of scatter in their relationship.

In the last row in Table~\ref{tab:FS_table}, we include the normalized ranges calculated from a set of uniformly distributed random numbers at the 4th ranking level. Comparing the ranges due to our features to these random ranges, we can tell that all four of our chosen features for all of our predicted quantities hold some information more than noise. Furthermore, from the relative ranges associated with these top four features, we can see that some of our predicted quantities may need more than four features to make accurate predictions, while others may be able to reach maximum accuracy with fewer features. For example, in the case of mass loss, the range due to the fourth feature, M\textsubscript{host}, is already small, and  closer to the range obtained by a uniform random number than in the case of, for instance, final position, where the range of the 4th feature remains relatively high, and well above noise.

Figure~\ref{fig:bestSpaces} shows predicted quantities as a function of the two most important features. For example, the top left panel shows the mean survival fraction as a function of both initial scale and subhalo to host mass ratio. Note the strong trends in predicted quantities in these 'best feature planes'. For survival, the plane is clearly divided, suggesting that the survival of a subhalo is already well-determined by only two features. For the other quantities, this gradient is less defined, suggesting that either more features are needed to make good predictions, or the process is stochastic enough that general trends with respect to our input features are harder to find. In particular, the panel for merge time (lower right) in Figure~\ref{fig:combined_distributions_logBin} shows little variation in outcome across the entire plane of initial scale and subhalo-to-host mass ratio, despite the merging time having the strongest trend of variation with those features. From these 'best feature planes', it is clear that some of these final quantities, such as the binary outcome of survival or disruption, will be much more straightforward to predict than others, such as merge time.


\subsection{Survival}
\label{sec:survival} 

\begin{figure}
	\includegraphics[width=\columnwidth]{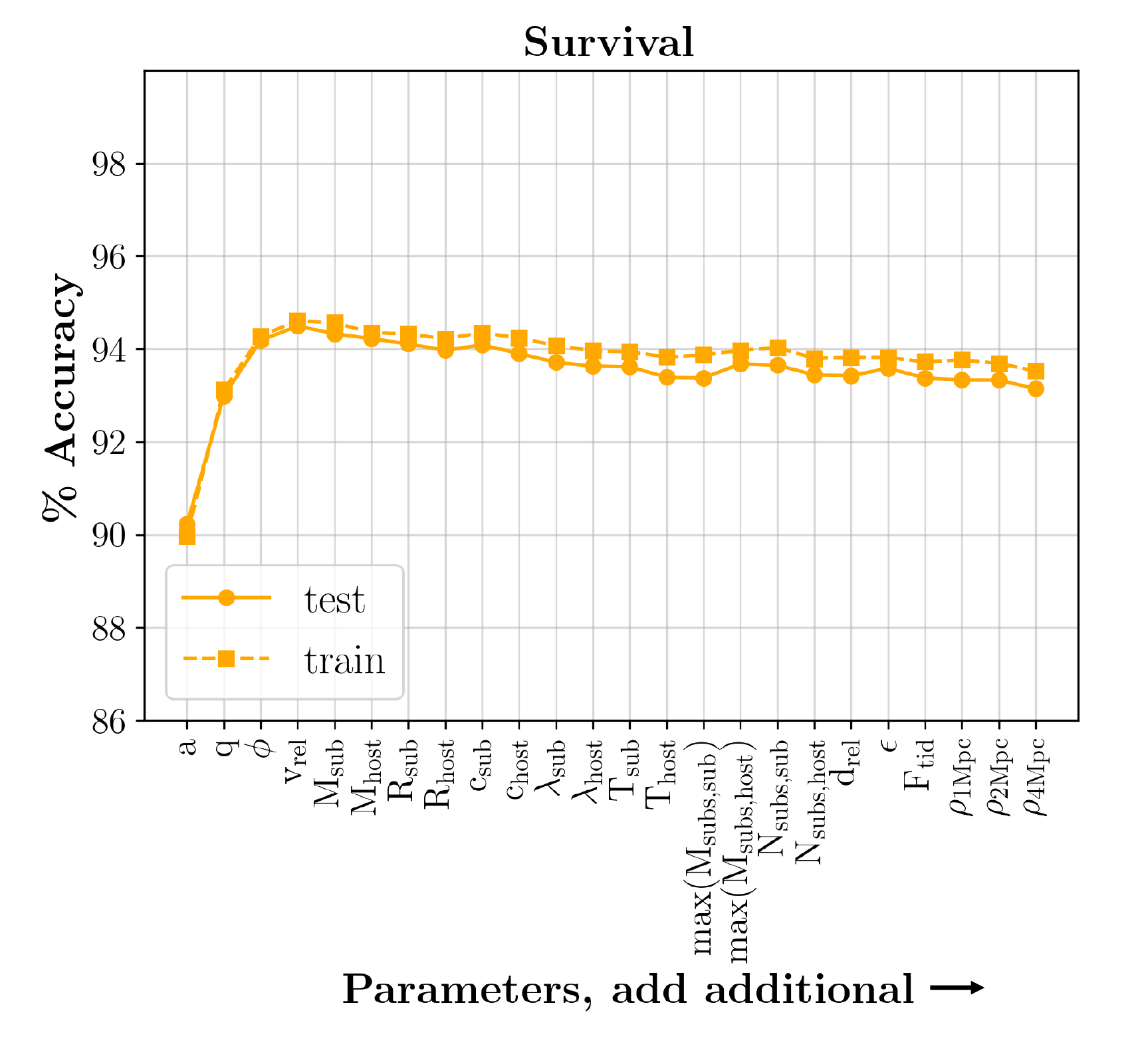}
	\vspace{-15pt}
    \caption{Accuracy of model predictions, for both the training and test sets, in the case of predicting subhalo survival. On the y-axis, we show the accuracy, defined as the percentage of the subhalo sample that is predicted correctly. On the x-axis, we show the features used to train the model. For each point, the model was trained using all features to the left of and including that point on the x-axis. The solid line and circles show the accuracy of the test set, while the dashed line and square points show the accuracy of the set the model was trained on. The choice and order of features for the first four features in the x-axis is determined by our feature selection algorithm described in Section~\ref{sec: param selection}, while the order of the remaining features is arbitrary.}
    \label{fig:survival_predictions}
\end{figure}

For the entire sample of subhalos, we predict whether or not a halo will survive until z=0 (assigned a 1) or dissolve within the host (assigned a 0) using a random forest classifier. The details of this model are outlined in Section~\ref{model details}. Our accuracy is defined  straightforwardly as the percentage of halos correctly classified. Figure~\ref{fig:survival_predictions} displays the accuracy when one feature at a time is added to the model in the order shown.  We emphasize that, although we add features to the model in an ordered way, the random forest does not use the ordering of input features when training a model; each new subset of features requires re-training the model. Thus, if the ordering of our features had been completely random, the maximum model accuracy would not change, although the slope of information gain in Figure~\ref{fig:survival_predictions} would.

As is clearly shown in Figure~\ref{fig:survival_predictions}, after adding the four most important features, we reach a maximum accuracy for both the testing and training sets of 94.4\% and 94.6\%, respectively. The gap between these accuracies is small, suggesting low overfitting. The decrease in accuracy (less than 1\%) when adding additional features beyond these four suggests a lack of information gained by adding any feature thereafter. Any small increase or decrease in the accuracy beyond the first four features are within noise. Since the model is free at any iteration to use as many of the provided features as needed, the fact that maximum accuracy is reached after the addition of only these four features suggests that they are the only features that are necessary to predict the survival of a subhalo.

\begin{figure}
	\includegraphics[width=\columnwidth]{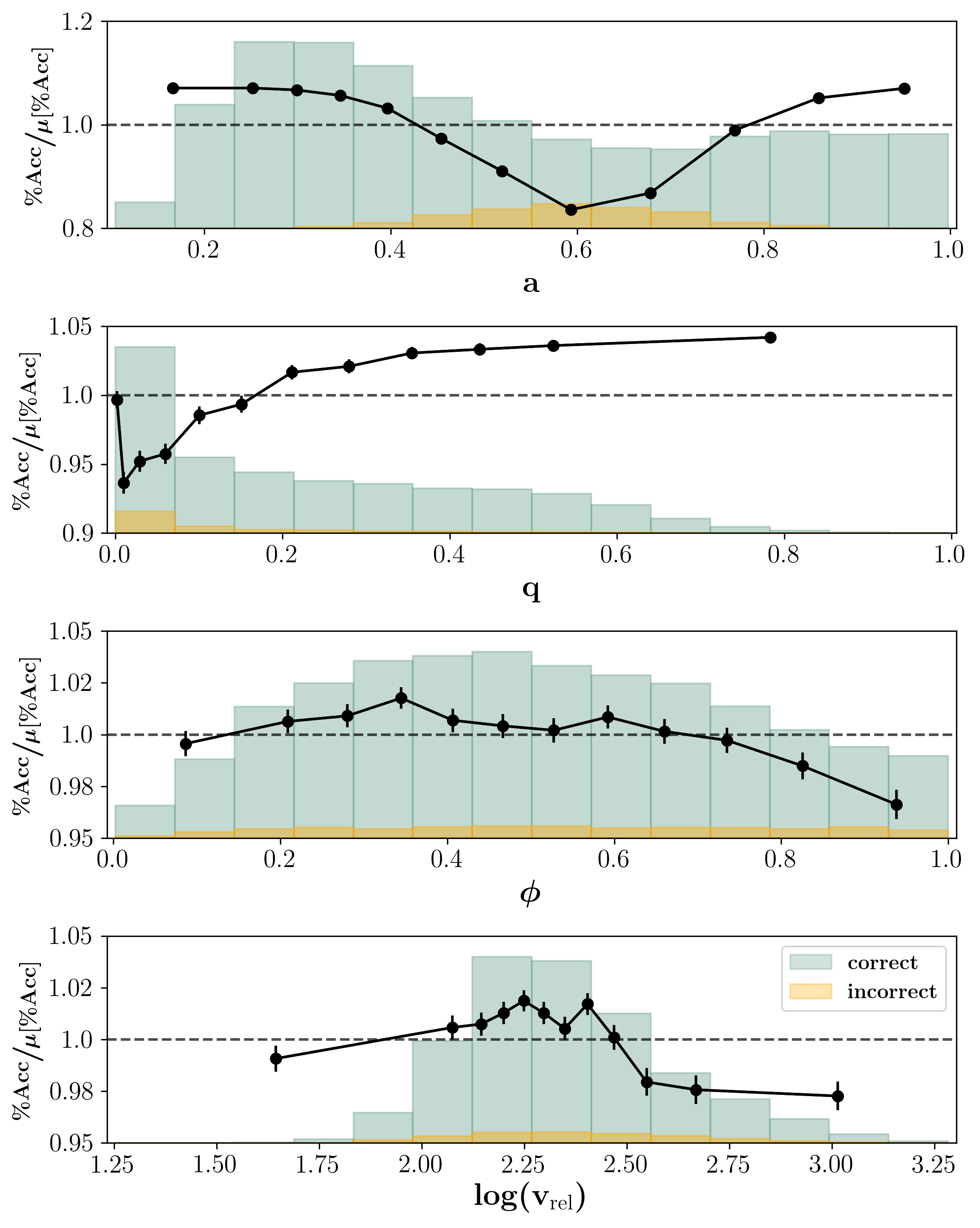}
	\vspace{-20pt}
    \caption{Accuracy of subhalo survival predictions for our test set, as a function of the four most important features, shown in four different panels. Black lines and points show the average percentage of accurate predictions within bins of each feature, normalized by the average percentage of accurate predictions of the entire test set. Bins along the x-axis are created such that the same number of subhalos belong to each bin. We include dashed lines where the y-axis value is 1. Above this line, predictions in that bin are on average better than the test set average, and below this line, predictions in the bin are on average worse than the test set average. The histograms show the distributions of the accurately (green) and inaccurately (orange) predicted populations. Since the survival quantity is binary, predictions can only either be correct (prediction matching truth) or incorrect (prediction not matching truth). We note that the histograms are normalized to the figure size and their height does not correspond to the y-axis labels.}
    \label{fig:surv_trends}
\end{figure}

The four critical features are:  $a$, $q$, $\phi$, and $v_\mathrm{rel}$. The initial scale of the subhalo entry is overwhelmingly the most important of these features - 90\% of our test sample is accurately predicted with this feature alone. With the addition of each of the remaining features, a 2.7\%, 1.2\%, and .27\% gain in accuracy occurs. In Figure~\ref{fig:surv_trends}, we show trends in accuracy with respect to each of these features, by showing the average accuracy score in bins, normalized by the average accuracy score of the entire sample. Here, we see distinct trends in accuracy with respect to each of our features. The initial scale of entry, which has the most predictive power for our sample, also has the most drastic trend with accuracy. All subhalos that enter their hosts at times either before $a$ = 0.3 or after $a$ = 0.9 are predicted correctly. This is consistent with what we see in the top left panel of Figure ~\ref{fig:combined_distributions_logBin}, where all subhalos entering after $a$ = 0.9 survive, and almost all halos entering before $a$ = 0.3 merge. Subhalos with entry times between $a$ = 0.5-0.7 are by far the hardest to make predictions for, with the peak of this uncertainty occurring at  $a$ = 0.6. There is also a strong trend with respect to $q$, where in general the larger the subhalo-to-host mass ratio is, the better survival is predicted. The one exception to this trend is at the smallest subhalo-to-host mass ratios, with $q$ < 0.002, where the percent of accurate predictions returns to about average. This is likely because most of the subhalos below this mass ratio do not survive, making them easier to predict. Subhalos on orbits with the highest $\phi$ (most grazing orbits) are worse predicted than those on orbits with $\phi$ < 0.7, and prediction accuracy appears to generally decrease for orbits that are both more plunging and more grazing than $\phi$ = 0.3-0.4. A similar trend occurs in log($v_\mathrm{rel}$), where subhalos with both larger and smaller relative velocities being worse predicted than subhalos with log($v_\mathrm{rel}$) = 2.1-2.4. For both $\phi$ and log($v_\mathrm{rel}$), the difference between the best and worst average accuracy points are much smaller than for $a$ or $q$, so these trends are also less significant.

In Figure~\ref{fig:surv_trends}, we also show the distributions of our correctly and incorrectly predicted subhalos. There is a clear tendency for subhalos with $a$ around 0.5-0.7 to be the most difficult to predict, corresponding to a range of redshifts of around z = 0.68-0.55. The distribution of correctly predicted subhalos peaks near $a$ = 0.3, or z = 2.3. As satellite occupation peaks at around z = 2.5 \citep{Wetzel2009}, this peak is likely due to most of the interactions occuring there. In fact, as the vast majority of our subhalos are correctly predicted, these distributions of the correctly predicted samples look nearly identical to the distributions of the complete sample. The peak of the incorrectly predicted subhalos also agrees with the region of greatest uncertainty that we see in the upper left plot of Figure~\ref{fig:bestSpaces}, where a clear division between always surviving and always dissolving occurs at $a$ =0.6. We note that, although most of the poorly predicted halos exist in this small section of feature space, 80.3\% of halos with $a$ =  0.5-0.65 are still accurately predicted, significantly better than random guessing. In the remaining panels of Figure~\ref{fig:surv_trends}, the distributions of the correctly and incorrectly predicted halos are roughly the same, and cover roughly the same range of values. In the bottom panel, the correctly predicted population peaks where the average accuracies are the highest, indicating a higher fraction of incorrect subhalos at low and high log($v_\mathrm{rel}$). 

To determine if there are significant differences between the correctly and incorrectly predicted subhalos with respect to our additional features beyond these most important four, we perform a KS test between the distributions of these two populations. We do this test using the distributions of correctly and incorrectly subhalos, with respect to each of the additional features beyond our set of the most important four. To ensure that any differences we find in these distributions are not due to correlations with our four most important features, we take a slice of our data in a narrow four-dimensional bin, that fixes a range of values for all of these four features. We then ensure that within this selected bin, the distributions of the correct and incorrect populations are the same according to the KS test. Then, we perform an individual KS test between the two distributions from this slice of data, with respect to each of our additional features. In doing so, we find that the two distributions are found to be the same, with a p-value of above 3\textsigma, for all of our additional features. This suggests that there is no significant difference in any of the additional features between correctly and incorrectly predicted subhalos.

\subsection{Mass Loss}
\label{sec:mass loss}
For all surviving subhalos at z=0, we predict the final mass using a gradient-boosting regressor, with hyperparameters as given in Table~\ref{tab:hyper_table}. To determine the accuracy of the model, we define an error metric, $\delta$(M), which we call the prediction error of each individual prediction, using the the difference between true and predicted fractional remaining mass. A subhalo is considered to be accurately predicted if:
\begin{equation}
    \label{eq:mass loss}
    \delta(M) = \frac{\lvert M_\textsubscript{pred,f} - M_\textsubscript{true,f}\rvert}{M_\textsubscript{true,i}} - \frac{2m\textsubscript{p}\sqrt{N\textsubscript{p,true,i}}}{M_\textsubscript{true,i}} \leq tol
\end{equation}
Where \textit{tol} is some tolerance value which determines what difference in fractional mass loss is acceptable as accurate. \textit{M} is the mass of the subhalo. $N_\mathrm{p}$ is the number of particles belonging to the subhalo, and $m_\mathrm{p}$ = 3.215 $\times$ 10\textsuperscript{7}h\textsuperscript{-1} M\textsubscript{\(\odot\)} is the mass of a dark matter particle in the simulation. Subscripts \textit{pred} refer to a value predicted by the model, while \textit{true} refer to the true value from the simulation. Subscripts \textit{f} denote quantities taken at z=0, and \textit{i} denote quantities taken when the subhalo first enters the host. We can then vary the tolerance, determining what percentage of subhalos have their prediction errors within certain tolerance thresholds. We point out that, because this prediction error is a measure of how close a prediction is to the truth, a higher prediction error value corresponds to a worse prediction, and a lower prediction error value means a better prediction.

\begin{figure}
	\includegraphics[width=\columnwidth]{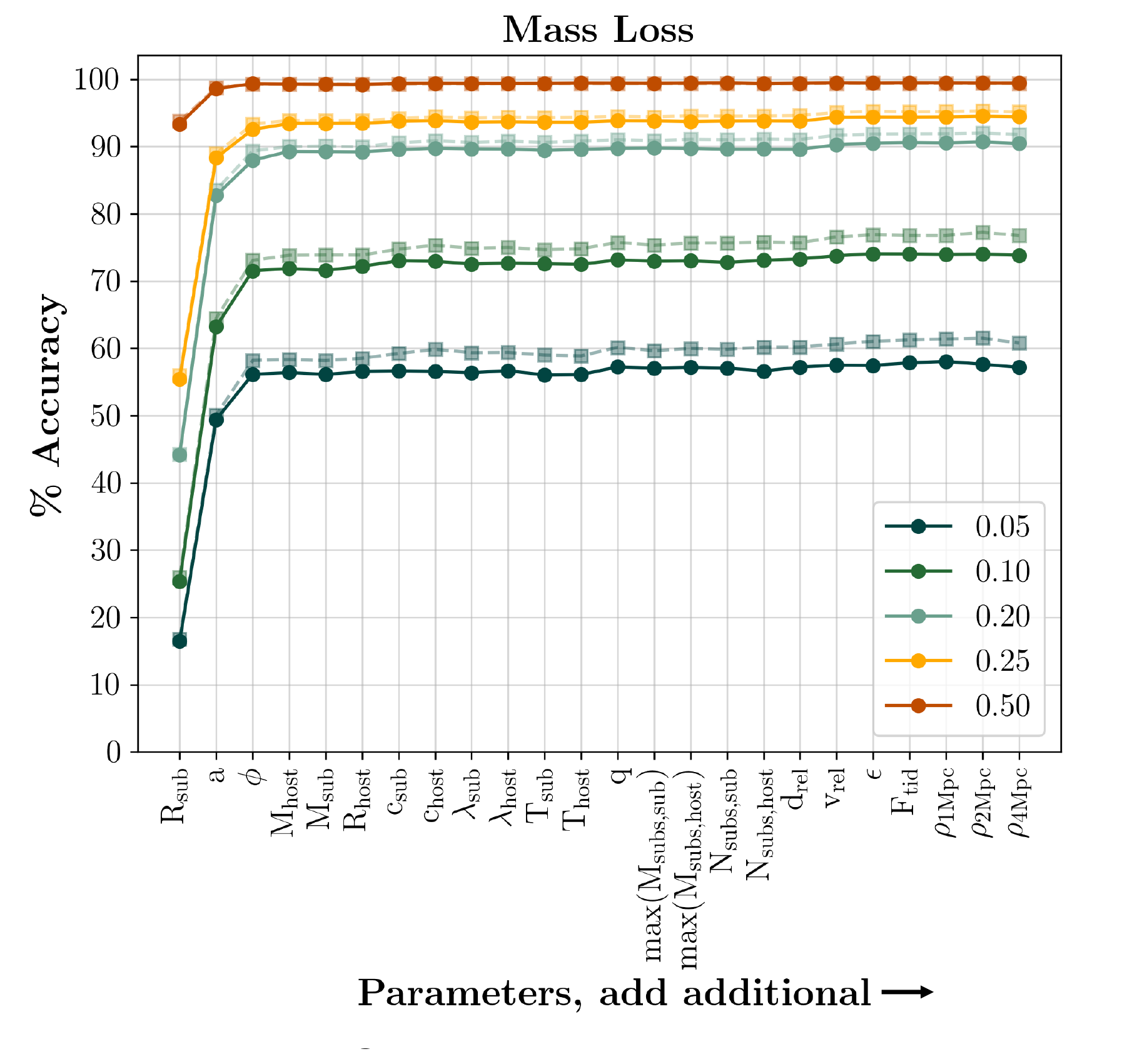}
	\vspace{-15pt}
    \caption{Accuracy of model predictions, for both the training and test sets, in the case of predicting subhalo mass loss. On the y-axis, we show the accuracy, defined as the percentage of the subhalo sample with a prediction error below some specified tolerance, as defined by Eq.~\ref{eq:mass loss}. On the x-axis, we show the features used to train the model. For each point, the model was trained using all features to the left of and including that point on the x-axis. The solid lines and circles show the accuracy on the test set, while the dashed lines and square points show the accuracy on the set the model was trained on. Different colored lines show the different tolerance values used to define accuracy. For a complete description of this accuracy metric, see the associated text. The choice and order of features for the first four features in the x-axis is determined by our feature selection algorithm described in Section~\ref{sec: param selection}, while the order of the remaining features is arbitrary.}
    \label{fig:massloss_predictions}
\end{figure}

The first term in this equation measures the difference between our true and predicted masses. Although the quantity that our machine learning model predicts is the mass of the subhalo, we determine prediction error by normalizing this value to the initial mass of the subhalo and comparing true and predicted fractions of initial mass. We do this in order to have a metric that equally penalizes errors in prediction for all subhalos, rather than allowing more leniency depending on the subhalo mass. The second term accounts for Poisson noise in the number of particles assigned to the subhalo. By subtracting this noise term from the error, we are stating that a prediction is perfect if it is within the Poisson noise limit. This term is significantly smaller than the first term and only makes a difference in the case of small subhalos where a small number of particles make up a large portion of the mass.
 
 \begin{figure}
	\includegraphics[width=\columnwidth]{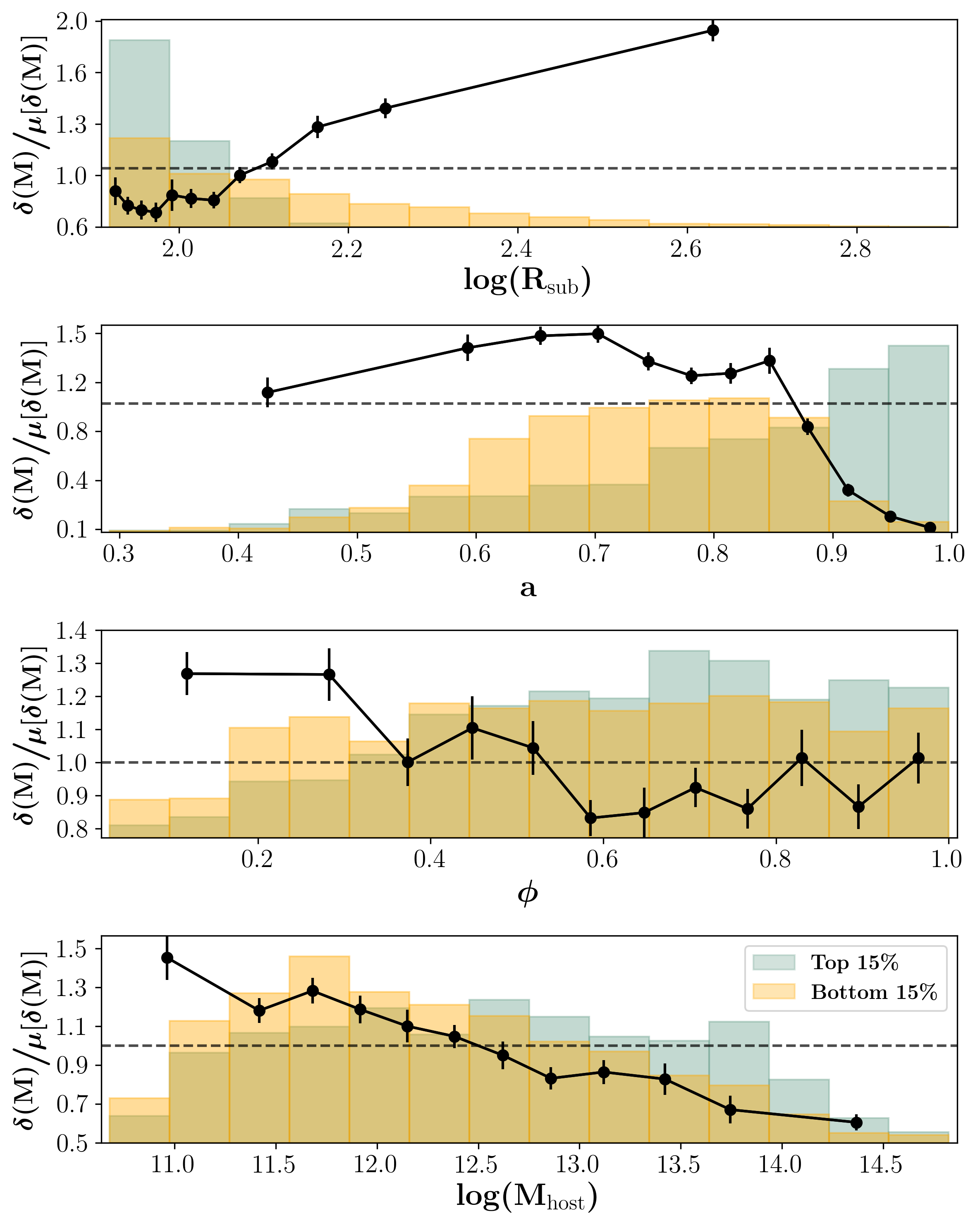}
	\vspace{-20pt}
    \caption{Prediction error for subhalo mass loss in our test set, as a function of the four most important features, shown in four different panels. Black lines and points show the average prediction error within bins of each feature, normalized by the average error of the entire test set. Bins along the x-axis are created such that the same number of subhalos belong to each bin. We include dashed lines where the y-axis value is 1. Our error metric is defined such that a lower value means a better prediction, so above this line, predictions in that bin are on average worse than the test set average, and below this line, predictions in the bin are on average better than the test set average. The histograms show the distributions of the 15\% best (lowest prediction errors; green) and 15\% worst (highest prediction errors; orange) predicted populations. We note that the histograms are normalized to the figure size and their height does not correspond to the y-axis labels. The error metric that we use for predicting subhalo mass loss is given by Eq.~\ref{eq:mass loss} and described in detail in the associated text.}
    \label{fig:mass_trends}
\end{figure}

Figure~\ref{fig:massloss_predictions} shows the accuracy of the model, when trained using the technique described above. Since the accuracy of this model depends on the selected tolerance value, we present our results for a range of tolerances. Again, the training set generally does better than the test set, for all tolerances, due to slight overfitting. It can be seen that, using only the four top-ranking features, 56.5\% of subhalos have their final masses accurately predicted to with a margin of error of less than \textpm 5\% of their true initial mass. 89.3\% of subhalos can be predicted accurately, given predictions within \textpm 20\% of their initial mass. Almost all (99.3\%) subhalos can have their masses predicted to within \textpm 50\% of their initial mass, although it's worth noting that this tolerance encompasses a very wide range of mass loss. 

To predict subhalo mass loss, the three most important features are: R\textsubscript{sub}, $a$, and $\phi$. Again, given the ordering from the feature selection method discussed previously, these features drive the steepest information gain, even given a model allowed to select any of the full 26 feature set. At the 20\% tolerance level, adding these first three features results in an increase of 44.2\%, 38.5\%, and 5.3\% accuracy percentage gain, respectively. We note that, because the radius and mass of subhalos are directly analytically related to one another, the radius can be replaced with the mass in this model with no difference in the information gain or final accuracy.

Figure~\ref{fig:mass_trends} shows the average prediction error, as a function of the four most important features for making these predictions. We create bins with respect to each of these 4 parameters, spaced such that that the same number of subhalos belong to each bin, and plot the average prediction error in that bin, normalized by the average prediction error of the entire sample. We also show distributions of the subhalos with the 15\% best and worst predicted mass loss. There are clear trends in the error of our predictions with respect to each of these four features. The prediction error increases with increasing R\textsubscript{sub}, so smaller subhalos are predicted better than larger subhalos. Prediction errors also decrease for subhalos that enter their hosts at more recent times. In particular, there is a sharp improvement to predictions for subhalos entering their hosts at $a$ $\geq$ 0.85. Subhalos entering their hosts around $a$ = 0.6-0.7 are the hardest to make predictions for. There is a slight trend in prediction error with regards to $\phi$. Subhalos that enter their hosts with $\phi$ $\leq$ 0.3 are slightly harder to make predictions for than those entering on more grazing (higher $\phi$) orbits. Finally, there is a roughly linear trend between prediction error and log(M\textsubscript{host}), with prediction error decreasing as log(M\textsubscript{host}) increases, meaning that it is easier to make predictions for subhalos entering larger hosts. This is likely related to mass ratio as well, as smaller hosts will tend to have interactions with larger $q$ than smaller hosts. Indeed, we find a similar trend in log($q$), with better predictions for subhalos with smaller mass ratios.

Interestingly, the distributions of the best and worst predicted subhalo populations do not always separate strongly. Most of the well-predicted subhalos are of smaller size, while the poorly predicted halos span a larger range of sizes, although their highest concentration is at a similar size to that of the well-predicted subhalos. We note that this is likely due to the fact that there are more small subhalos than large, so the majority of our sample falls within this range. However, the distribution of poorly predicted subhalos does tell us that, despite the trend in prediction error with subhalo size, many of our smaller subhalos are still difficult to make predictions for. As the radius of a subhalo is analog to its mass, this also means that less massive subhalos are better predicted than their more massive counterparts. The well-predicted subhalos also tend to reside in larger host masses than their poorly predicted counterparts, suggesting that the well-predicted population is more comprised of unequal mass ratios. The best-predicted subhalos are also those that enter their host at later times, with the contours centering around $a$ = 0.9-0.95, likely because those do not have much time to lose mass before the end of the simulation, and thus have final masses similar to their initial masses. The most concentrated regions of poorly predicted subhalos also trace the regions of highest prediction error well. The best predicted subhalos appear to have slightly higher impact angles than their poorly predicted counterparts, although the total span is roughly the same for both.

As before, we want to determine if there are significant differences between the best and worst predicted subhalos with respect to our additional features. As we did with the correctly and incorrectly predicted populations for our survival predictions, we perform a KS test between these distributions. This is done with respect to each of the additional features beyond our set of the most important four, after finding a narrow bin within these four features that removes all differences between the best and worst predicted halos with respect to those four features. The KS test shows that the two distributions are found to be the same, with a p-value of above 3\textsigma, for all of our additional features, meaning that no  additional feature exhibits a trend with the goodness of our predictions.

\subsection{Final Position}
\label{sec:position}
We next predict the final position of a surviving subhalo at z=0, relative to the center of the host, using a gradient-boosting regressor. To determine the accuracy of the model, we define a positional error metric, $\delta$($d_\mathrm{rel,f}$), which we call the prediction error for each of our predictions, using the the difference between true and predicted fractional distance from host center. A subhalo is considered to be accurately predicted if:
\begin{equation}
    \label{eq:position}
    \delta(d\textsubscript{rel,f}) = \lvert d_\textsubscript{rel,f,true} - d_\textsubscript{rel,f,pred}\rvert - \frac{2R\textsubscript{soft}}{R_\textsubscript{host,f}} \leq tol
\end{equation}
Where \textit{tol} is some tolerance value that determines what difference in fractional distance from host center is acceptable as accurate. Here, $d_\mathrm{rel,f}$ is the distance between subhalo and host halo centers, normalized by the host radius. Subscripts \textit{pred} refer to a value predicted by the model, and \textit{true} refer to the true value from the simulation. $R\mathrm{host,f}$ is the radius of the host at z=0, and R\textsubscript{soft} is the softening length of the simulation. The first term in this equation is simply the absolute difference between the true and predicted fractional distance from host center. Since what our model predicts is the actual fractional distance of the subhalo from the host halo center, we do not need to additionally normalize this quantity as we did when predicting mass loss, as this value can be straightforwardly taken as the fraction of the host radius by which the prediction is off. The second term in the equation is an additional tolerance, to account for uncertainty in the subhalo's position within its host due to the force resolution of the simulation. As with the mass loss predictions, we vary the tolerance to determine what percentage of subhalos have their prediction errors within that tolerance, which is how we define the model accuracy.

\begin{figure}
	\includegraphics[width=\columnwidth]{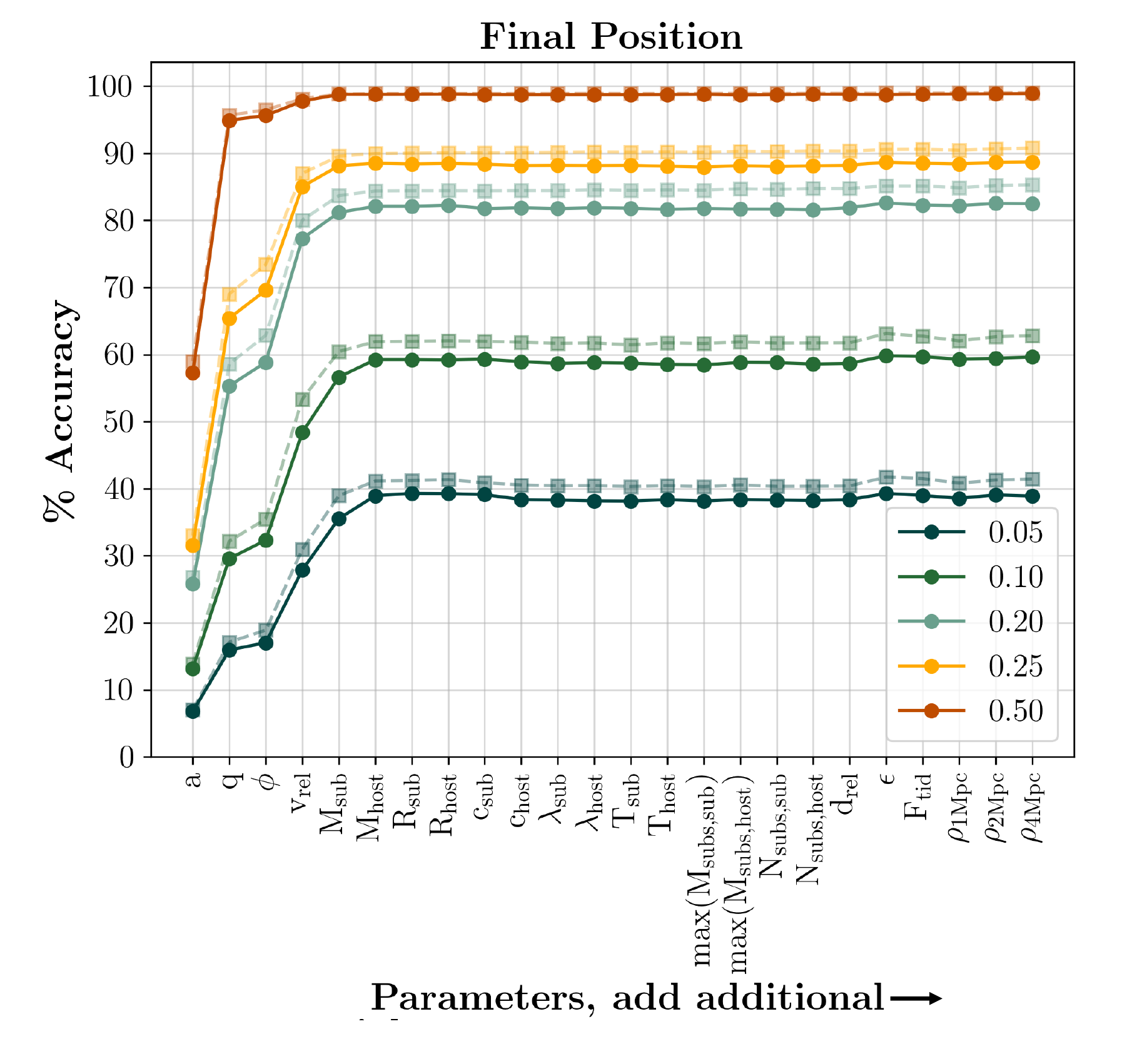}
	\vspace{-15pt}
    \caption{Same as Fig.~\ref{fig:massloss_predictions}, but for the case of predicting the final subhalo position. Different colored lines show the different tolerance values used to define accuracy, as defined in Eq.~\ref{eq:position}. For a complete description of the positional error metric we use to calculate this accuracy, see the associated text. The choice and order of features for the first four features in the x-axis is determined by our feature selection algorithm described in Section~\ref{sec: param selection}, while the next two features were chosen by the GBR algorithm. The order of the remaining features is arbitrary.}
    \label{fig:position_predictions}
\end{figure}

Figure~\ref{fig:position_predictions} shows the accuracy of the model. For final subhalo position, it appears that more features are needed to reach maximum accuracy, with the first six required before accuracy converges. Moreover, the fraction of well-predicted halos is smaller than in the case of predicting mass loss. 39\% of subhalos have their final positions accurately predicted to \textpm 5\% of their final host radius, 82.1\% of subhalos can be predicted accurately to within \textpm 20\% of their host's radius, and 98.8\% to within \textpm 50\%. 

\begin{figure}
	\includegraphics[width=\columnwidth]{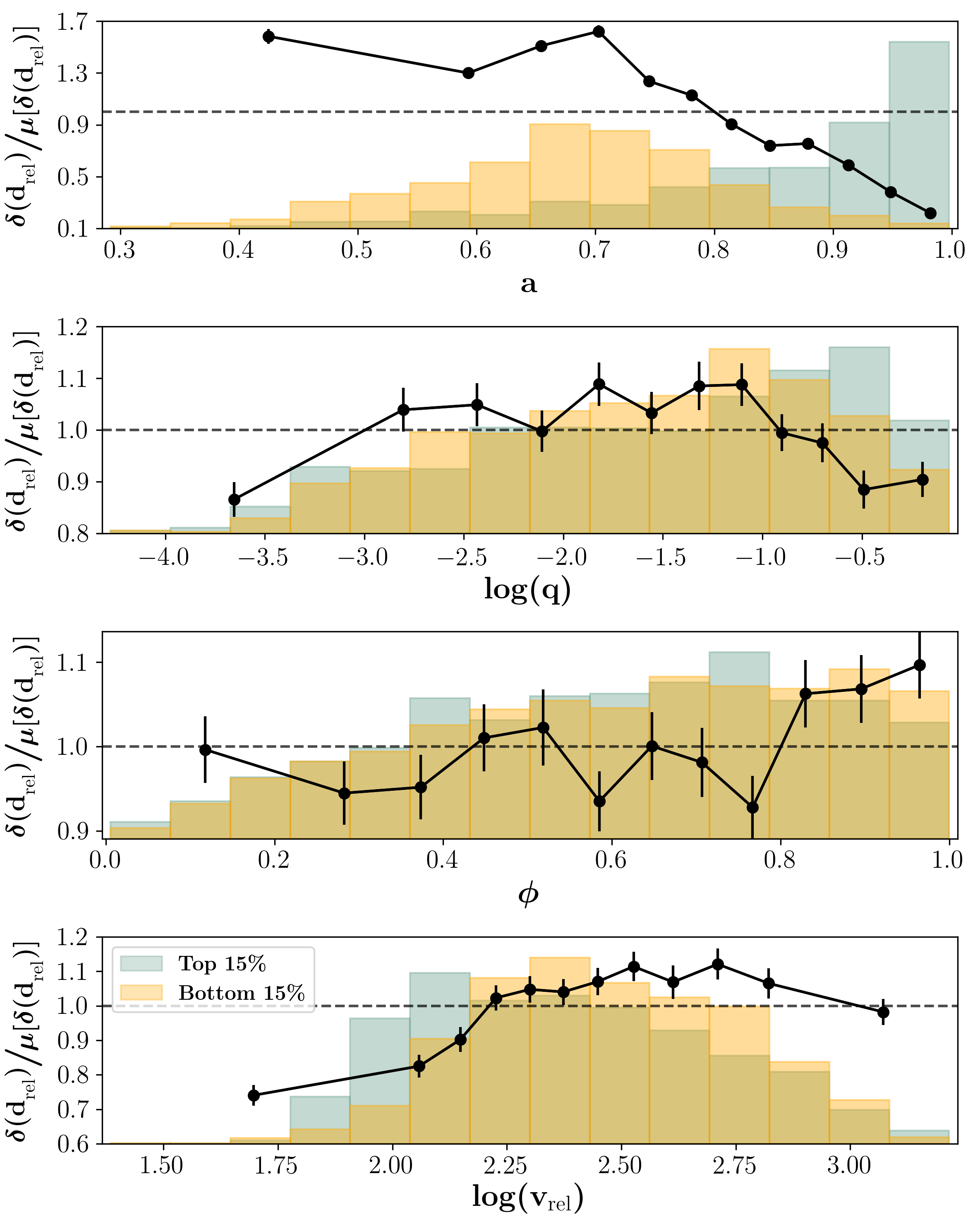}
	\vspace{-20pt}
    \caption{Same as Fig.~\ref{fig:mass_trends}, but for the case of predicting the final subhalo position. The error metric that we use for predicting subhalo position is given by Eq.~\ref{eq:position} and described in detail in the associated text. This error metric is also defined such that lower values on the y-axis correspond to better predictions.}
    \label{fig:pos_trends}
\end{figure}

The most important six features to predict final position are: the initial scale factor, the mass ratio between the sub and host halo, the subhalo's orbital impact angle, the relative velocity with which the subhalo enters, the mass of the subhalo, and the mass of the host halo. Given the ordering from the feature selection method discussed previously, it appears that these first four features were chosen to be quite important, although the subhalo impact angle provides less accuracy gain than some of the other, later-chosen features. The additional two features that we did not find with our feature selection methods, the mass of the subhalo and the mass of the host halo, were found by the machine learning model to be additionally important. Adding these first six features results in an increase of 25.8\%, 29.6\%, 3.6\%, 18.3\%, 3.9\%, and 1\% accuracy percentage gain, respectively, at the 0.2 tolerance level. In this case, our feature selection method does not add information in the optimal order, so some later added features provide more information gain than earlier selected features. For each of our four top features, we show trends in prediction error in Figure~\ref{fig:pos_trends}. There is a strong trend with regards to $a$, where the later a subhalo enters its host, the better its final position can be predicted. As with mass loss, this is likely because subhalos entering their hosts closest to z=0 have less time to undergo significant changes from the influence of their host, or fall very deeply into the host center. There is a significant decrease in prediction error for subhalos entering later than $a$ = 0.7, with subhalos entering prior to that time being generally predicted poorly. Subhalos with both lower mass ratios, log($q$) < -3, and higher mass ratios, log($q$) > -1, are predicted better than those at more mid-range mass ratios, with prediction error notably decreasing for the higher mass ratios.  There appears to be little trend in prediction error with regards to $\phi$. Higher impact angles ($\phi$ > 0.8) are predicted slightly worse than subhalos on more plunging orbits, but below this impact angle, there is no significant trend. Similarly, subhalos with log($v_\mathrm{rel}$) < 2.2 are predicted better than subhalos with larger initial velocities, but above this initial velocity there appears to be little trend. This is perhaps due to subhalos incoming with the smallest initial velocities not changing position significantly from their time of entry, making them easier to make predictions for.

Figure~\ref{fig:pos_trends} also shows distributions of where the best and worst 15\% of predicted subhalos lie. Most of the best-predicted subhalos are those that enter their host closer to z=0, likely because those have less time to move deep into the host and have their orbits altered. The distribution of poorly predicted subhalos peaks at an earlier time, around $a$ = 0.7, as those subhalos likely spend more time in their host halos with the potential for larger, less predictable perturbations. Well-predicted subhalos also seem to slightly favor more equal mass ratios than poorly-predicted ones, but the difference between the distributions is fairly minor, except for at log($q$) > -1.0, or mass ratios of greater than 1:10, where the higher relative number of well-predicted subhalos to poorly predicted subhalos brings the average prediction error down. From the third panel of Figure~\ref{fig:pos_trends}, it appears that best and worst distributions with the impact angle are quite similar. There are slightly fewer poorly predicted subhalos than well predicted subhalos at more grazing orbits with $\phi$ > 0.8, which likely explains the increase in average prediction error at those values. Well-predicted subhalos favor slightly lower initial velocities than their poorly-predicted counterparts, with the peak of the well-predicted population occurring at log($v_\mathrm{rel}$) = 2.1, and the peak of the poorly-predicted population being closer to log($v_\mathrm{rel}$) = 2.3.

As before, we check the distributions of the best and worst predicted subhalos with respect to each of our additional features, beyond our set of the most important four, using a KS test. After controlling for $a$, $\phi$, $q$, and $v_{rel}$ by selecting a narrow bin in this feature space where the two distributions of best and worst predicted subhalos are the same, we perform a KS test between the two distributions with respect to all additional features. In doing so, we find that the two distributions are the same, with a p-value of above 3\textsigma, for all of our additional features, except for the concentration of the subhalo and the eccentricity.

\subsection{Merge Time}
\label{sec:merge time}
For those subhalos that dissolve before z=0, we predict the time between the subhalos entry and subsequent merging, using a gradient boosting regressor. To determine the accuracy of the model, we define an merger time error metric, $\delta$(t), which again we refer to as the prediction error for an individual subhalo, using the the difference between true and predicted number of crossing times. A subhalo is considered to be accurately predicted if:
\begin{equation}
    \label{eq:time}
    \delta(t) = \frac{\lvert t_\textsubscript{true} - t_\textsubscript{pred}\rvert}{t_\textsubscript{cross,true}} \leq tol
\end{equation}
Where \textit{tol} is some tolerance value which determines to within how many crossing times a prediction is considered to be accurate. Here, \textit{t} is the predicted duration of the merger, and $t_\mathrm{cross}$ is the crossing time of the host halo, at the time that the subhalo dissolves, both in years. Subscripts \textit{pred} refer to a value predicted by the model, and \textit{true} refer to the true value from the simulation. Then, our tolerance is in units of final crossing times of the host halo. This $\delta$(t) is calculated as the difference between the true and predicted elapsed time of infall for the subhalo, normalized by its final crossing time. Normalizing by this crossing time allows us to use this error metric for all subhalos, regardless of when the interaction occurs. As before, by varying the tolerance, we calculate the percentage of halos with acceptable prediction errors to get accuracy. 

\begin{figure}
	\includegraphics[width=\columnwidth]{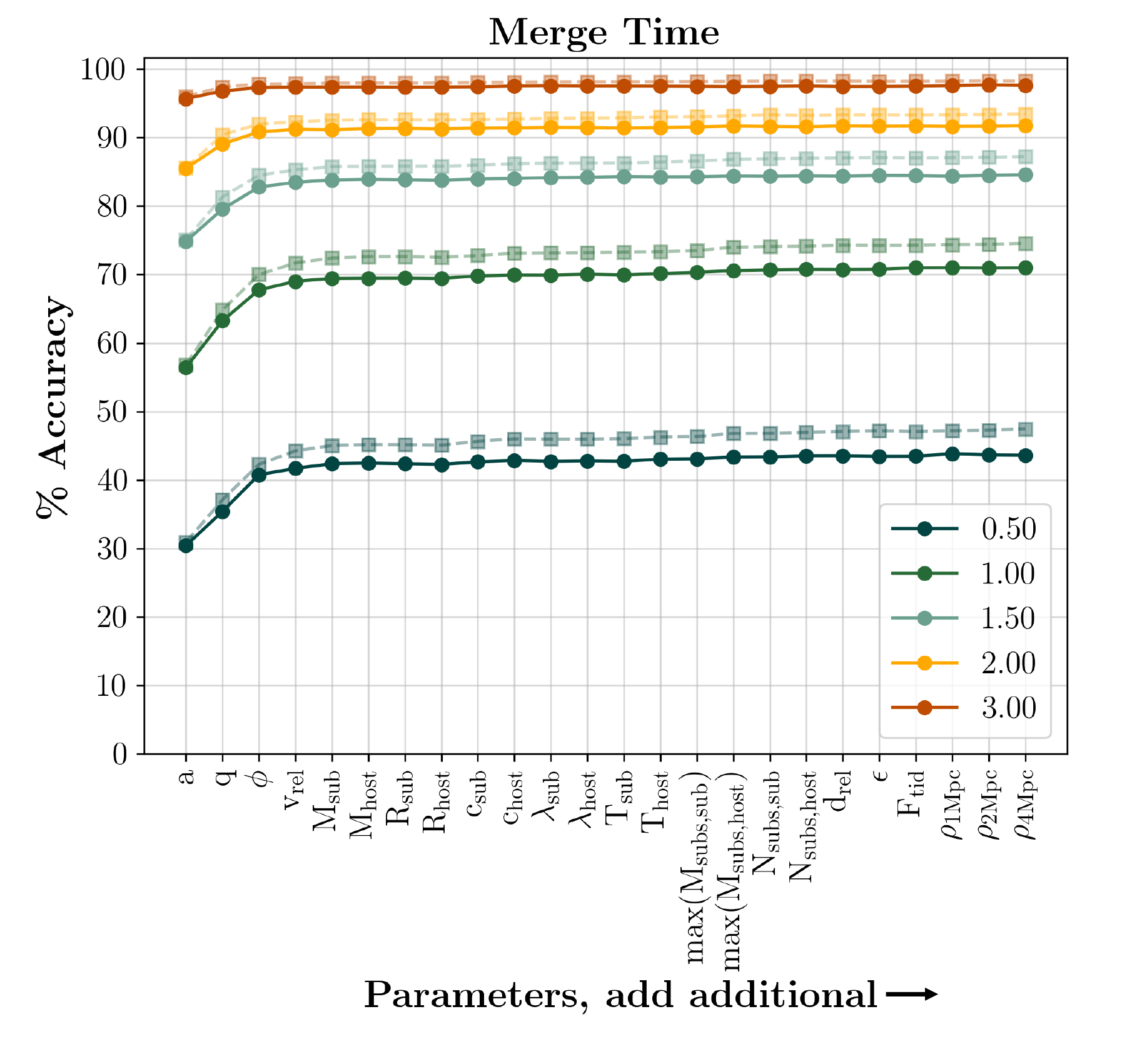}
	\vspace{-15pt}
    \caption{Same as Fig.~\ref{fig:massloss_predictions}, but for the case of predicting the subhalo merge time. Different colored lines show the different tolerance values used to define accuracy, as defined in Eq.~\ref{eq:time}. For a complete description of the error metric we use to calculate this accuracy, see the associated text.}
    \label{fig:time_predictions}
\end{figure}

Figure~\ref{fig:time_predictions} shows the accuracy of the model, when trained using all and increasingly smaller subsets of the features. Since accuracy depends on the selected tolerance value, we show the accuracy given several different choices of tolerance. As always, the training set generally does better than the test set, for all tolerances, due to slight  overfitting. To predict subhalo merging time, only three features appear to be necessary to reach maximum accuracy. 41.6\% of subhalos can be predicted to within half of a crossing time, 83.5\% of subhalos can be predicted to within 1.5 crossing times, and 97.4\% can have their merging time predicted to within 3 crossing times. We note that 3 crossing times is typically a few billion years, and is around the average time it takes a subhalo to merge, so this threshold is very lenient.

To test if this performance is an indication of our machine learning model learning any complex behavior, we compare the results of our model to a baseline model that assigns a constant merging time to each subhalo equal the average number of crossing times of all subhalos to merge. Then, we use Equation~\ref{eq:time} to check the accuracy of this simple model and compare it to the accuracy of our machine learning model. We find that, using this baseline model, 40.3\% of subhalos are correctly predicted to within .5 crossing times, 83\% of subhalos are correctly predicted to within 1.5 crossing times, and 98.8\% of subhalos are correctly predicted to within 3 crossing times. The performance of this simple model is extremely similar to that of our machine learning model, at all tolerance levels. A similar comparison for our other predicted quantities showed that those machine learning models performed much better than a baseline model.

\begin{figure}
	\includegraphics[width=\columnwidth]{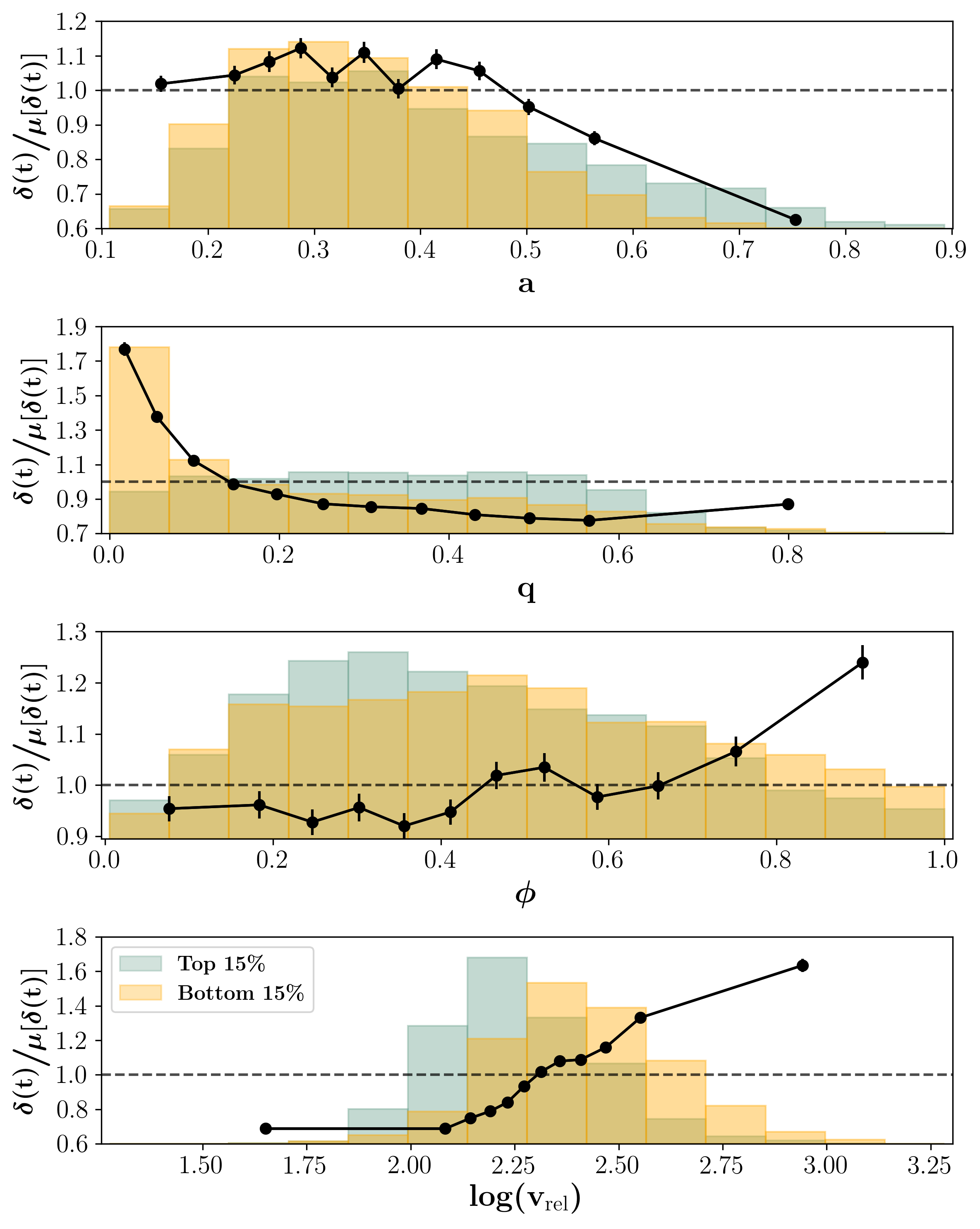}
    \vspace{-20pt}
    \caption{Same as Fig.~\ref{fig:mass_trends}, but for the case of predicting the subhalo merge time. The error metric that we use for predicting subhalo merge time is given by Eq.~\ref{eq:time} and described in detail in the associated text. This error metric is also defined such that lower values on the y-axis correspond to better predictions.}
    \label{fig:time_trends}
\end{figure}

The three features needed before prediction accuracy levels off with the addition of more features are: the initial scale factor, the mass ratio between the sub and host halo, and the subhalo's orbital impact angle. Adding these first three features results in an increase of 74.9\%, 4.7\%, and 3.1\% percentage gain in accuracy, respectively, at the 1.5 tolerance level. In Figure~\ref{fig:time_trends}, we show trends in prediction error with respect to each of the top four features. In the top panel, we see a trend with $a$ for subhalos with entry times at $a$ > 0.45, where later entry times are on average predicted better. For subhalos entering at times earlier than $a$ = 0.45, there does not appear to be a trend. Subhalos with larger $q$ are also better predicted than those at more unequal mass ratios, with the prediction error rapidly decreasing as mass ratios become more equal, until around $q$ = 0.3, where the trend roughly levels off. This mass ratio of $q$ = 0.3 is often presented as the threshold between major and minor mergers \citep{Wetzel2009}, so major mergers are predicted much more easily than minor mergers. This trend is likely due to the fact that major mergers happen more quickly than minor mergers do, and the more equal the mass ratio in a merger is, the more quickly the merger occurs. As such, we would expect higher \textit{q} mergers to be easier to predict because the target value is smaller. Subhalos on more grazing initial orbits, with $\phi$ > 0.7 have on average higher prediction errors than those on more plunging orbits. Plunging orbits likely take less time to merge than grazing orbits, and thus likely also have less time for their orbits to be changed significantly, making them easier to predict. Initial relative velocity has a consistent trend with prediction error, where the higher log($v_\mathrm{rel}$) is, the worse a subhalo is predicted.

Figure~\ref{fig:time_trends} also shows the distributions the subhalos with the 15\% best and worst prediction errors. The poorly predicted subhalos have a slight tendency to enter their hosts at earlier times than their better predicted counterparts, however the majority of both distributions are subhalos entering their hosts at earlier times. Most of the best predicted subhalos have more equal mass ratios, whereas the worst predicted subhalos are highly concentrated in the smallest $q$ subhalos. The distribution of the best predicted subhalos peaks at more plunging $\phi$ orbits than the worst-predicted subhalos, with a higher concentration of poorly predicted subhalos at higher $\phi$, following the trends that we saw with prediction error. Finally, the distributions in log($v_\mathrm{rel}$) look similar but offset, with the best-predicted subhalo distribution having a peak at around log($v_\mathrm{rel}$) = 2.2, and the worst-predicted subhalo distribution having a peak at around log($v_\mathrm{rel}$) = 2.3. This offset is small, but consistent with the trends in prediction error that we see.

We check the distributions of the best and worst predicted subhalos with respect to each of our additional features, beyond our set of the most important four. After controlling for a, $\phi$, $q$, and $v_{rel}$ by selecting a narrow feature space where the two distributions of best and worst predicted subhalos are the same, we perform a KS test between the two distributions with respect to all additional features. In doing so, we find that the two distributions are the same, with a p-value of above 3\textsigma, for all features except for the spin of the host halo and the concentration of the host halo.

\subsection{Subhalo Interactions}
\label{sec:interactions}

Several papers have noted that interactions between subhalos as they orbit within their hosts can be frequent and lead to significant amounts of mass loss, with as much as 40\% of mass loss in a subhalo attributed to subhalo interactions~\citep{Tormen1998, Knebe2005,Klimentowski2010, Angulo2009}. In our sample, we find that interactions between  subhalos are quite common. Around 57\% of our subhalos spend at least one snapshot as a sub-subhalo; that is, they enter the radius of another subhalo within the host at some point after they have entered the host itself. After becoming a sub-subhalo, around 23.5\% of our total sample remain shrouded as sub-subhalos until they dissolve or until z = 0.

To test if these close interactions are important in determining mass loss, or any of our other predicted final quanitities, we track and incorporate three additional subhalo features in our model: (1) a flag indicating whether the subhalo becomes a sub-subhalo; (2) a flag indicating whether the subhalo remains a sub-subhalo until either merging or z=0; and (3) the time that the subhalo spends being a sub-subhalo. Since the subhalo can enter and then exit another subhalo multiple times, this number may reflect time spent inside more than one subhalo.  Although these features are tracked during the whole history of a subhalo's infall, and thus do not align with our initial goal of predicting outcomes of subhalos using only initial conditions, we add these additional features to our model solely to determine if they matter significantly. We find that none of these features increase accuracy when added, meaning that the number and duration of interactions does not inform the evolution of our subhalo final quantities. We also tested whether these interaction features could add enough stochasticity to the merging process to be responsible for the difficulty in making these predictions by checking if the distributions of the best and worst predicted subhalos are significantly different. For each model, we control for the features that were found to be important by selecting a narrow bin within each of them, then we perform a KS test between the distributions of best and worst predicted subhalos with respect to our interaction features. In doing so, we find that the two distributions are the same, with a p-value of above 3\textsigma, for all of our models. This suggests that these close interactions between subhalos are not important for their evolution as they fall into their hosts, neither by affecting the outcome nor by adding noise to the process. 

\section{Summary and Discussion}
\label{sec:Conclusion}
In this paper, we employed machine learning algorithms to predict the survival, mass loss, final position, and merge time of a subhalo from features taken at the time of its initial infall into its host halo. Our goal was to better understand to what degree these final outcomes are due to stochasticity in subhalo evolution versus real, physically-motivated processes that could be consistently, analytically predicted. we found:
\begin{itemize}
    \item Subhalo survival can be predicted remarkably well, with 94.4\% of our sample being correctly predicted as surviving or disrupting. To reach this accuracy, four initial features are needed: the scale factor at the time of the start of the interaction, the mass ratio between the subhalo and its host, the impact angle of the subhalo's orbit, and the initial relative velocity between the subhalo and its host. However, to reach this accuracy, the initial scale factor is by far the most influential of these features, and an accuracy of 89.9\% can be reached with this feature alone. Subhalos with both late and early entry times are easiest to predict, while those entering their host halos at $a$ = 0.6 are more difficult. However, this is also dependent on the subhalo-to-host mass ratio, where subhalos with lower mass ratios instead exhibit this transition closer to  $a$ = 0.3. This is likely because lower mass ratio subhalos are in general more likely to survive, so a subhalo must enter its host at an earlier time to be subject to changes from its host for long enough to dissolve.
    \item Subhalo mass loss is a much more stochastic process. Although for 56.5\% of our sample we were able to predict a final mass with an error within \textpm5\% of the initial mass, we must loosen our criteria to \textpm20\% of the initial mass in order to consider $\sim$ 90\% of our sample correctly predicted. This maximum prediction accuracy is achieved using only three initial features: the radius of the subhalo, the scale factor at the time of the start of the interaction, and the impact angle of the subhalo orbit. In general, our model makes better predictions for smaller subhalos with late entry times than for those that are larger or have earlier infall times.
    \item Subhalo final positions are also difficult to predict. 39\% of our sample can be correctly predicted to within \textpm5\% of their host's initial radius, but an accuracy of 88.5\% is only achieved when we loosen our error tolerance to within \textpm25\% of the host radius. To make these predictions, six initial features are needed: the scale factor at the time of the start of the interaction, the mass ratio between the sub and host halo, the impact angle, the initial relative velocity, the subhalo mass, and the host halo mass. As with mass loss, our model makes better predictions for subhalos entering their hosts at later times. However, there does not seem to be a significant trend in prediction accuracy with regards to the other features that were found to be important for making these predictions. 
    \item Subhalo merging timescales are also difficult to predict. 41.9\% of our sample can be correctly predicted to within half of their host halo's final crossing time, but an accuracy of 91.1\% is only achieved when we loosen our error tolerance to within 2 crossing times. Our model needs four features to make its predictions: the scale factor at the time of the start of the interaction, the mass ratio between the sub and host halo, the impact angle, and the initial relative velocity. Notably, this model shows no improvement in accuracy over a naive model which assigns to all subhalos the average number of crossing times our subhalos take to merge.
    \item There are some interesting commonalities among both the sets of features needed to make these predictions and the feature spaces in which predictions are poorest. Only five features, in total, are needed to achieve the maximum prediction accuracy for all of our predicted outcomes:. The scale factor, impact angle, relative velocity, and the masses of the host and subhalo (sometimes combined as mass ratio or appearing as virial radius instead) seem to be the only relevant features for determining subhalo evolution. Additionally, the feature spaces that are most difficult to make predictions within also have much overlap. In general, subhalos that enter at a mid-range of initial scales (typically a = 0.6-0.7) are challenging to make predictions for, across all of our final outcomes. There also appear to be trends in impact angle, with higher impact angles (more grazing orbits) being easier to predict the behavior of, except for in the case of predicting disruption, where the opposite appears to be true.
    \item Additional features beyond the set needed to make predictions for each final quantity are not useful, either for making predictions or for characterizing the types of subhalos that are better or worse predicted. Although the best and worst predicted subhalos are typically distributed differently with respect to the features that are used to make predictions, when these features are controlled for, differences in these distributions with respect to all other features are removed. So, additional features outside of the set used to make predictions do not correlate with the stochasticity of our predictions.
\end{itemize}

It is clear from our results that, for predicting the mass loss, final location, and merging timescales of individual subhalos, an accurate, consistent mapping for a significant fraction of the population cannot be found given our set of initial features. There are several possible reasons for this inability to accurately model subhalo evolution. The first possibility is that some feature or features were missing from the initial set, which would have been fundamental to making accurate predictions. Although we have made sure to include an extensive list of physically-motivated features that were found in the literature to be important for modeling these outcomes, our list was not completely comprehensive. For instance, the halo finder ROCKSTAR outputs 75 features to describe each subhalo, many of which encode information about the ID's of the halos, but also include: halfmass radius, largest shape ellipsoid axes, angular momenta, velocity dispersion, and some others, most of  which we decided not to include in our analysis. However, we have no compelling reason to believe that these excluded features would contribute such a meaningful portion of the needed information to bridge this gap in predictability. So, although the possibility remains that additional features could be needed to improve predictions, it seems unlikely that the missing information could be completely encompassed there.

A second possibility is that errors within the simulation, halo catalog, or merger tree make subhalo evolution unpredictable and sometimes incorrect. Much speculation remains as to the accuracy of N-body simulations and their ability to accurately model the physics of subhalo evolution, particularly on these small scales~\citep{Kampen2005, Taylor2005, VandenBosch2018, Bosch2018}. Although several studies have suggested that simulation resolution only effects subhalos with small numbers of particles~\citep[e.g.,][]{Gao2004, Nurmi2006, Diemand2007}, \citet{VandenBosch2018} found that typical state-of-the-art cosmological simulations cannot resolve subhalos well enough to follow their mass loss until complete disruption. In the Bolshoi simulation, for example, \citet{VandenBosch2017} found that only around 20\% of subhalo disruption was truly physical, with instantaneous subhalo masses being highly erratic along the orbit. Similarly, \citet{VandenBosch2018} found that most subhalo disruption in modern simulations is artificial or numerical in nature, with only subhalos of exquisite resolution and greater than 10\textsuperscript{6} particles per halo showing consistently converged results, especially for those orbiting close to the center of the host. A number of other works have also called into question the reliability of the halo catalogs and merger trees that are generated from these simulations. Comparison projects have found differing results for the fates of subhalos and the subhalo mass functions resulting from different halo finders~\citep{Knebe2011, Onions2012, Avila2014, Bosch2014, Behroozi2015} and merger tree codes~\citep{Tweed2009, Srisawat2013, Jiang2014}. Because these codes fundamentally define subhalos in different ways and trace their properties between snapshots using different methods, these comparison projects found that, when applied to the same simulation, resulting halo catalogs and merger trees could differ quite significantly. This problem may be additionally exacerbated by the frequently tumultuous merger histories of halos \citep{Sinha2012}.

Despite the fact that the veracity of simulation results has been brought into question, we note that it seems unlikely that simulation errors could be entirely responsible for the results we have found. \citet{Bosch2018} found that, despite physical disruption\footnote{We mean "physical" here as it is used in \citet{Bosch2018}, where it refers to disruption caused by the tidal heating and stripping that unbounds the subhalo particles, as opposed to numerical disruption, which is due to a subhalo falling below the resolution limit of the simulation. In the case of numerical disruption, the subhalo would still exist if the resolution of the simulation were higher.} being extremely rare within simulations, subhalos only became highly sensitive to numerical disruption after losing over about 90\% of their mass. Additionally, \citet{Avila2014} and \citet{Srisawat2013} found that spurious fluctuations in the masses of subhalos within simulations using ROCKSTAR may be frequent, and that subhalos that pass close to the centers of their hosts may have truncated merger trees. However, they also found that Consistent-Trees is usually able to successfully follow the evolution of subhalos, leading to overall reliable mass loss histories. Since our study only relies on final outcomes mapping from the initial conditions, any minor errors along the history should not be important. And, although different halo finders and merger tree codes can differ from one another, their behavior is generally consistent, and thus again should not be responsible for the inconsistent merging behavior that we have found.

Another possibility is that final outcomes can be accurately predicted from initial conditions of the interaction, but the machine learning methods used here were not able to capture the process. This could be due to a few reasons. It is possible that the particular machine learning method used was not well-suited to the problem. However, we sampled different algorithms, from very simple methods to more complicated methods such as neural networks, and have found that this technique yields the highest fraction of accurate predictions. Another possibility is that we did not have enough data, or the data did not span the parameter space well enough. In this instance, noise in the data could overwhelm the relationship between input and output. We also tested this by repeating our process with smaller random subsets of our data, using instead one half and one quarter of our original sample, to ensure that our results did not change with fewer data points. In doing this, we found no change in maximum accuracy, even with a significantly reduced amount of data, suggesting that the noise in the data is not due only to sample size.

The final possibility is that there is an inherent chaotic nature of these interactions within N-body simulations. We find a large scatter in outcomes to be present in our data - even in narrow bins of the input parameters, there can be large differences in the outcome with regard to all of our predicted quantities, meaning that a consistent relationship between these inputs and outputs can not be found. As we do not expect the previous possible explanations to completely explain this behavior, we believe that a chaotic nature of these interactions is the most likely explanation. In this case, the initial conditions of an interaction are not enough to know the outcome, because similar initial conditions can lead to very different outcomes. This makes it impossible for a model that relies on similar initial conditions producing similar outcomes to get consistent results.

An inherent stochasticity in these merging processes would have implications for simulations and the models that are built from them. For instance, our findings in this work would suggest that analytic models that attempt to model the individual evolution of subhalos are doomed to be unable to describe all subhalo-host halo interactions, as the basic premise that the same inputs would yield the same outputs is not necessarily true. However, the regions of highest uncertainty in our predictions may give insights on how to improve these models. For instance, in predicting merge time, our model clearly had more trouble making predictions for minor mergers over major mergers. This could point to a need to model the merging times of these two populations separately, as our machine learning model, which is able to perform well in the major merger regime, clearly does not apply as well to the minor merger regime. In this instance, there may be an inconsistent dependence of merging timescales on the mass ratio, making it difficult to accurately parameterize its effect.

Despite the inability of these models to make accurate predictions at the individual subhalo level, we may still be able to use a model like this to construct subhalo populations. For instance, to determine the surviving population of subhalos at z=0 given a population of subhalos that enter the host, our model would be able to definitively determine the survival or disruption of some subhalos that enter within certain ranges of our most important features, and determine with some probability the survival or disruption of other subhalos that enter within the more uncertain ranges of our features. A model like this could properly take into account the regions of feature space where outcomes are more variable, to assign outcomes with some scatter, but model more precisely in the regions of feature space where outcomes are better defined. In this way, one could use our model to create realistic distributions of z=0 subhalo populations using moderate resolution simulations that do not actually resolve subhalo evolution. 

The features that our model selects also give some interesting insights. For instance, predicting the survival, final position, and merge time of a subhalo, all required the same four features: $a$, $q$, $\phi$, and $v_\mathrm{rel}$, all in the same relative order of importance, to make predictions. This likely means that the same fundamental physical processes are at work in determining all of these quantities. To predict mass loss of a subhalo, our model selected from the same broad subset of features, but the features that provided the most information to the model were not the same. This may mean that some additional processes effect subhalo mass loss, that skew some features to be more necessary in making predictions than were needed for the other final quantities. Another interesting result from the features that our model selected is that the impact angle and relative entry velocity of a subhalo as individual features appear to be more influential in determining subhalo evolution than a feature like eccentricity, which captures information about both. This may mean that allowing a model to individually weigh those two components of subhalo orbits leads to a more generalizable model of subhalo evolution than using eccentricity or circularity alone. 

Finally, as our analysis here has shown us trends in the outcomes of subhalos with respect to certain features, we can use this information to draw conclusions about satellite populations. For instance, the overwhelming dependence of survival on the entry time of the subhalo, along with the clear division that we see in Figure~\ref{fig:bestSpaces} of survival fraction occurring at z = 0.67-0.43, suggests that all satellite galaxies that we see today must have entered their host halo more recently than these times. Alternatively, due to the trend we also see in mass ratio, they may only have entered at an earlier time if they entered their host with a mass ratio of less than 1:100. Similarly, we could expect a roughly smooth trend in the fraction of halo mass remaining with the time of entry for subhalos.

In this study, we used a dark matter only simulation to study the evolution of subhalos. However, hydrodynamic simulations play a crucial role in our understanding of subhalo evolution by capturing a more complete context of baryonic physics. The next step in this type of work would naturally be to explore the implications of baryons. Because there are more potential factors dictating the evolution of subhalos in hydrodynamic simulations, such as feedback and enhanced tidal effects \citep{Diemand2011, Brooks2013, Despali2017}, we may expect that subhalo evolution in a hydrodyanmic simulation is even more difficult to predict. A recent study by \citet{Nadler2017} perhaps confirms this, as they used dark matter properties to predict the survival of subhalos in a hydrodyanmic simulation, with slightly less success than we were able to predict survival in a dark matter only simulation, meaning that the addition of baryonic physics makes this prediction more difficult. However, whether or not adding baryonic features to these types of models would improve predictions remains unexplored. Several works have studied the specific effects of baryons on the subhalos in which the galaxies reside~\citep{Dolag2009, Romano-Diaz2010, Brooks2014, Sawala2016, Garrison-Kimmel2017, Munshi2017, Richings2018}, but as can be seen from this work, the chaotic nature of these interactions may make it difficult to quantify these trends into a machine learning model.

\section*{Acknowledgements}

A.A.B. was supported in part by the National Science Foundation (NSF) through a Career Award (AST-1151650). This research made use of several software packages, including NumPy \citep{van2011numpy}, SciPy \citep{Virtanen_2020}, Scikit-learn \citep{scikit-learn}, seaborn \citep{seaborn}, pandas \citep{McKinney_2010, McKinney_2011} and matplotlib, a Python library for publication quality graphics \citep{Hunter2007}. This work was conducted in part using the resources of the Advanced Computing Center for Research and Education at Vanderbilt University, Nashville, TN. This work used the Extreme Science and Engineering Discovery Environment (XSEDE) \citep{xsede}, which is supported by National Science Foundation grant number ACI-1548562, through allocation TG-AST130037. Parts of this research were conducted by the Australian Research Council Centre of Excellence for All Sky Astrophysics in 3 Dimensions (ASTRO 3D), through project number CE170100013.




\bibliographystyle{mnras}
\bibliography{mendeley.bib}




\bsp	
\label{lastpage}
\end{document}